\documentclass[a4paper,11pt]{article}

\usepackage{jheppub} 

\usepackage[T1]{fontenc} 
\usepackage{bm,latexsym,amsmath,amssymb,amsfonts,mathtools,caption}
\usepackage{amssymb, amsmath, comment}
\usepackage{color}
\usepackage{hyperref}
\usepackage{booktabs}
\usepackage{mathrsfs}
\usepackage{ulem}
\usepackage{wrapfig}
\makeatletter
\newcommand{\figcaption}[1]{\def\@captype{figure}\caption{#1}}
\newcommand{\tblcaption}[1]{\def\@captype{table}\caption{#1}}

\newcommand{\kdsf}{Kerr-AdS${}_5$ }
\makeatother
\preprint{RIKEN-iTHEMS-Report-22}
\begin{document}
\title
{Global study of the scalar quasi-normal modes of Kerr-AdS${}_5$ black holes:\\
Stability, thermality, and horizon area quantization}
\author[a]{Issei Koga}
\author[b]{Naritaka Oshita}
\author[a]{Kazushige Ueda}
\affiliation[a]{Department of Physics, Kyushu University, 744 Motooka, Nishi-Ku, Fukuoka 819-0395, Japan}
\affiliation[b]{RIKEN iTHEMS, Wako, Saitama 351-0198, Japan}

\abstract{
We numerically explore the structure of quasi-normal (QN) frequencies of the five-dimensional small and large Kerr-anti de Sitter (Kerr-AdS${}_5$) black hole with equal and unequal rotations. Our investigation also covers low and high Hawking temperatures. We then study the stability of the \kdsf black hole and the structure of highly damped QN modes, which would reflect the thermodynamic property of the \kdsf black hole. We find that the highly damped complex QN frequencies of a nearly maximally spinning \kdsf black hole have the periodic separation of the surface gravity at the horizon in the imaginary part while the real part converges to the superradiant frequency, which may be relevant to the pole structure of the thermal Green's function in the corresponding conformal field theory on the \kdsf boundary. Finally, we discuss a relation between the QN modes of the \kdsf black hole and the Hod's conjecture on the horizon area quantization along with the analysis of the horizon topology of the \kdsf black hole. We show that in general, an ultra-spinning \kdsf black hole, whose spin parameter is infinitesimally close to the AdS curvature radius, has its non-compact horizon, and based on the Hod's conjecture, we argue that the horizon area may be continuous, that is, the unit area of the horizon vanishes in the ultra-spinning regime.}
 
\maketitle

\section{Introduction}
Higher-dimensional gravity is important in the holographic principle \cite{tHooft:1993dmi,Susskind:1993if,Susskind:1994vu,Maldacena:1997re}, the brane-world scenario \cite{Horava:1996ma,Arkani-Hamed:1998jmv,Randall:1999vf,Randall:1999ee,Arkani-Hamed:1999wga}, and the landscape \cite{Susskind:2003kw} of the string theory. In these contexts, black holes in higher dimensions play pivotal roles, e.g., thermality of the conformal field theory (CFT) on the anti-de Sitter (AdS) boundary \cite{Witten:1998qj,Banks:1998dd}, the source of dark radiation in brane-world models \cite{Kraus:1999it,Ida:1999ui,Mukohyama:1999qx}, and so on. 
Another interesting nature of higher-dimensional black holes is that there is no corresponding theorem to a uniqueness theorem in higher dimensions, although it exists in the four-dimensional spacetime. Thus the stability of black holes in higher dimensions is non-trivial. The analysis of quasi-normal (QN) modes of black holes is a good probe to see its stability. 
The positivity of the imaginary part of complex QN frequencies leads to the exponential growth of the perturbation of black holes. The structure of QN modes is also relevant to the context of the holography as the QN modes of black holes in AdS space are conjectured to be dual with the poles of Green's function of the CFT on the AdS boundary (see Ref. \cite{Birmingham:2001pj} for the simplest case of the duality in the BTZ black hole). The structure of QN modes of black holes could shed light on the quantum nature of black hole horizons as well. According to the Hod's conjecture \cite{Hod:1998vk}, the real part of highly damped QN frequencies determines the unit area of a horizon. It would be an interesting question if the concept of the horizon area quantization can be extended to higher-dimensional gravity for which there are rich structures such as black holes, black rings, black strings, and so on.

In this work, we consider the QN frequencies of a scalar field in the \kdsf spacetime \cite{Myers:1986un} to understand the stability, thermality, and area quantization of the black hole horizon. The scalar perturbations of the \kdsf black hole are governed by the Klein-Gordon equation that reduces to the Heun's differential equation by performing the coordinate transformations and the redefinition of the perturbation variables \cite{Aliev:2008yk}. Therefore, the solution of the equation is represented by the Heun function.

The analysis of AdS${}_5$ spacetime with a black hole and/or compactified extra-dimensional space has been done mainly for holographic applications \cite{Murata:2008xr,Murata:2009jt,Cardoso:2013pza,Arefeva:2020jvo,Nian:2020qsk}. Analytical and numerical studies of the QN modes of the \kdsf spacetime have been recently done in Ref. \cite{BarraganAmado:2018zpa,Amado:2021erf,daCunha:2021jkm} by performing the expansion of the $\tau$-function for the Painlev\'{e} transcendents. In special cases, e.g., nearly equal spins, small spins, near extremal, or small mass regimes, the expansion of the $\tau$-function converges fast and makes the computation of QN frequencies tractable. On the other hand, in our computation, we use the solution of the Teukolsky equations represented by the Heun functions, which makes the computation of the QN modes substantially fast, and allows us to investigate broader parameter regions of the \kdsf spacetime. Also, our computation can be applied to search for not only the fundamental QN mode but also other overtones, including highly damped modes. Our computation with those advantages allows us to make a ``global map'' of the locations of QN modes of \kdsf spacetime, which would be useful to understand the stability of the background spacetime and even the Kerr-AdS/CFT correspondence \cite{Amado:2017kao}.

In the next section, we provide a review of the scalar perturbation of the \kdsf black hole. In Sec. \ref{sec:equal}, we study the QN modes of the \kdsf black hole with equal spins in the small and large mass regimes. In Sec. \ref{sec:unequal}, we investigate the QN modes for the unequal spins in the small and large mass regimes. In Sec. \ref{sec:areaQ}, we discuss some implications to the area quantization of the \kdsf black hole by applying the Hod's conjecture to our results. Also, we analyze the topology of the horizon of the \kdsf black hole in the ultra-spinning limit for which its spin parameter is infinitesimally close to the AdS curvature radius $\ell$. Finally, we summarize our results and conclusions in Sec. \ref{sec:conclusion}.

\section{Formalism}
\label{sec:formalism}
The \kdsf spacetime has the following metric
\begin{align}
\begin{split}
ds^2 &= - \frac{\Delta_r}{\rho^2} \left( dt - \frac{a_1 \sin^2 \theta}{1-a_1^2} d \phi - \frac{a_2 \cos^2\theta}{1-a_2^2} d \psi \right)^2 + \frac{\Delta_{\theta} \sin^2 \theta}{\rho^2} \left( a_1 dt - \frac{r^2+a_1^2}{1-a_1^2} d\phi \right)^2\\
&+ \frac{1+r^2}{r^2 \rho^2} \left( a_1 a_2 dt - \frac{a_2 (r^2 +a_1^2) \sin^2 \theta}{1-a_1^2} d\phi - \frac{a_1 (r^2 + a_2^2) \cos^2 \theta}{1-a_2^2} d\psi \right)^2 \\
&+ \frac{\Delta_{\theta} \cos^2\theta}{\rho^2} \left( a_2 dt - \frac{r^2+a_2^2}{1-a_2^2} d\psi \right)^2 + \frac{\rho^2}{\Delta_r} dr^2 + \frac{\rho^2}{\Delta_{\theta}} d\theta^2,
\end{split}
\label{metric}
\end{align}
where $M$ is the mass parameter, the AdS curvature radius is set to $\ell= 1$, $a_1$ and $a_2$ are the spin parameters for the two rotations of the \kdsf black hole, and
\begin{align}
\Delta_r &\equiv \frac{1}{r^2} (r^2+a_1^2) (r^2+a_2^2) (1+r^2) -2M,\\
\Delta_{\theta} &\equiv 1-a_1^2 \cos^2 \theta -a_2^2 \sin^2 \theta,\\
\rho^2 &\equiv r^2 + a_1^2 \cos^2 \theta+a_2^2 \sin^2 \theta.
\end{align}
Based on the thermodynamic description of the \kdsf spacetime, the Arnowitt-Deser-Misner (ADM) mass and angular momentum are given by \cite{Gibbons:2004ai,Hollands:2005wt,Olea:2006vd}
\begin{equation}
{\cal M} \equiv \frac{\pi M (2 \Xi_1 + 2\Xi_2-\Xi_1 \Xi_2)}{4 \Xi_1^2 \Xi_2^2}, \ 
{\cal J}_{\phi} \equiv \frac{\pi M a_1}{2 \Xi_1^2 \Xi_2}, \ 
{\cal J}_{\psi} \equiv \frac{\pi M a_2}{2 \Xi_1 \Xi_2^2},
\label{ADM_variables}
\end{equation}
where $\Xi_i \equiv 1-a_i^2$ ($i=1,2$). The spin parameters are restricted to $a_i \leq 1$, for which all the physical quantities in (\ref{ADM_variables}) are well-defined.  We here compute the QN modes of the \kdsf black hole and investigate the instability of a scalar field $\Phi (t,r,\theta, \phi, \psi)$ with mass $\mu$. Let us start with the Klein-Gordon equation
\begin{equation}
\left[\Box -\mu^2 \right] \Phi = 0,
\end{equation}
and decomposing $\Psi$ as $\Psi = e^{-i \omega t +im_1 \phi + i m_2 \psi} \Theta (\theta)\Pi (r)$, one has the radial and angular equations:
\begin{align}
\begin{split}
& \frac{1}{r} \frac{d}{dr} \left( r \Delta_r \frac{d \Pi (r)}{dr} \right)  - \left[ \lambda + \mu^2 r^2 + \frac{1}{r^2} \left( a_1 a_2 \omega -a_2 (1-a_1^2)m_1 - a_1 (1-a_2^2) m_2 \right)^2  \right] \Pi (r)\\
&+ \frac{(r^2 + a_1^2)^2 (r^2 + a_2^2)^2}{r^4 \Delta_r} \left( \omega - \frac{m_1 a_1 (1-a_1^2)}{r^2+a_1^2} - \frac{m_2 a_2 (1-a_2^2)}{r^2+a_2^2}  \right)^2 \Pi (r) = 0,
\end{split}\\
\begin{split}
&\frac{1}{\sin \theta \cos \theta} \frac{d}{d\theta} \left( \sin \theta \cos \theta \Delta_{\theta} \frac{d \Theta (\theta)}{d\theta} \right) - \left[-\lambda + \omega^2 + \frac{(1- a_1^2) m_1^2}{\sin^2 \theta} + \frac{(1- a_2^2) m_2^2}{\cos^2 \theta} \right.\\
&\left. - \frac{(1-a_1^2) (1-a_2^2)}{\Delta_{\theta}} (\omega+m_1 a_1+m_2 a_2)^2 + \mu^2 (a_1^2 \cos^2\theta +a_2^2 \sin^2 \theta) \right] \Theta (\theta) =0,
\end{split}
\end{align}
where $\lambda$ is the separation constant to be determined so that $\Theta (\theta)$ is regular at $\theta = 0$ and $\theta = \pi/2$ \cite{Aliev:2008yk}.
Performing the following transformations:
\begin{align}
r &\to z \equiv \frac{r^2- r_-^2}{r^2 - r_0^2},\\
\Pi(r) &\to R(z) \equiv z^{\theta_-/2} (z-z_0) ^{\theta_+/2} (z-1)^{-\Delta/2} \Pi(z),\\
\sin^2 \theta &\to u \equiv \frac{\sin^2 \theta}{\sin^2 \theta - \chi_0}, \ \text{with} \ \chi_0 \equiv \frac{1-a_1^2}{a_2^2 -a_1^2},\\
\Theta(\theta) &\to S (u) \equiv u^{-m_1/2} (u-1)^{-\Delta/2} (u-u_0)^{-m_2/2} \Theta (u),
\end{align}
with $r_-$ and $r_+$ being the inner and outer horizon radii, respectively, $r_0$ being the imaginary root of $\Delta_r$, $\Delta \equiv 2 + \sqrt{4+ \mu^2}$, and $u_0 \equiv (a_2^2 -a_1^2)/(a_2^2 -1)$, the radial and angular equations reduce to the Heun's differential equations:
\begin{align}
&\frac{d^2 R}{dz^2} + \left[ \frac{1- \theta_-}{z} + \frac{-1 + \Delta}{z-1} + \frac{1- \theta_+}{z-z_0}  \right] \frac{d R}{dz} + \left( \frac{\kappa_1 \kappa_2}{z (z-1)} - \frac{K}{z (z-1) (z-z_0)} \right) R = 0,\label{heun_radial}\\
&\frac{d^2 S}{du^2} + \left[ \frac{1+ m_1}{u} + \frac{-1+\Delta}{u-1} + \frac{1 + m_2}{u-u_0}  \right] \frac{d S}{du} + \left( \frac{q_1 q_2}{u (u-1)} - \frac{Q}{u (u-1) (u-u_0)} \right) S = 0,
\label{heun_angular}
\end{align}
where
\begin{align}
\theta_i &\equiv \frac{i}{2 \pi} \frac{\omega - m_1 \Omega_{i,1} -m_2 \Omega_{i,2}}{T_i},\\
T_i &\equiv \frac{r_i^2 \Delta_r'(r_i)}{4 \pi (r_i^2+a_1^2) (r_i^2 +a_2^2)}, \ 
\Omega_{i,1} \equiv \frac{a_1 \Xi_1}{r_i^2 +a_1^2}, \ \Omega_{i,2} \equiv \frac{a_2 \Xi_2}{r_i^2 +a_2^2},\\
\kappa_1 &\equiv - \frac{1}{2} \left( \theta_- + \theta_+ -\Delta - \theta_0 \right), \ 
\kappa_2 \equiv - \frac{1}{2} \left( \theta_- + \theta_+ -\Delta + \theta_0 \right),\\
q_1 &\equiv \frac{1}{2} \left( m_1 + m_2 +\Delta - \zeta \right), \ 
q_2 \equiv \frac{1}{2} \left( m_1 + m_2 +\Delta + \zeta \right),\\
\zeta &\equiv \omega + a_1 m_1 + a_2 m_2,\\
\begin{split}
K &\equiv -\frac{1}{4} \left\{ \frac{\lambda + \mu^2 r_-^2 - \omega^2}{r_+^2 - r_0^2} + (z_0-1) [(\theta_+ + \theta_- -1)^2 -\theta_0^2-1] \right.\\
&~~~~~~~~~~~~~~~~~~~~~~~~~~~~~~~~~~~~~~~~~~~~~~~~~\left. +z_0 \left[ 2 (\theta_+-1) (1- \Delta) + (2-\Delta)^2 -2 \right] \right. \biggr\},
\end{split}\\
\begin{split}
Q &\equiv - \frac{1}{4} \left\{ \frac{\omega^2 + a_1^2 \mu^2 - \lambda}{a_2^2 -1} +  u_0 \left[ (m_2+\Delta -1)^2 -m_2^2 -1 \right] \right.\\
&~~~~~~~~~~~~~~~~~~~~~~~~~~~~~~~~~~~~~~~~~~~~~~~~~\left. + (u_0-1) \left[ (m_1+m_2+1)^2 -\zeta^2-1 \right] \right. \biggr\}.
\end{split}
\end{align}
The general solution of the Heun's differential equation, (\ref{heun_radial}), is
\begin{align}
\begin{split}
R &= c_0 R_{\rm in}(\omega, z) +d_0 R_{\rm out} (\omega,z)\\
&\equiv c_0 H \ell \left( \frac{z_0}{z_0-1}, \frac{-K}{z_0-1}; \kappa_1, \kappa_2, 1-\theta_+, \Delta-1; \frac{z_0-z}{z_0-1} \right)\\
& +d_0 \left( \frac{z_0-z}{z_0-1} \right)^{1-\epsilon} H\ell \left( \frac{z_0}{z_0-1}, \frac{\theta_+ [z_0 (\Delta -\theta_-) -1+\theta_-]}{z_0-1} -\frac{K}{z_0-1}; \right.\\
&\left. ~~~~~~~~~~~~~~~~~~~~~~~~~~~~~~~~~\kappa_1+\theta_+, \kappa_2 + \theta_+,
1+\theta_+, \Delta-1; \frac{z_0-z}{z_0-1} \right),
\end{split}
\end{align}
for $z \sim z_0 (r \sim r_+)$, and
\begin{align}
\begin{split}
R &= c_1 R_{\rm AdS} (\omega,z) + d_1 R_{\rm Div} (\omega, z)\\
&\equiv c_1 H\ell \left( 1-z_0, \kappa_1 \kappa_2 -\tilde{K}; \kappa_1, \kappa_2, \delta-1, 1-\theta_-; 1-z \right)\\
&+d_1 (1-z)^{\theta_-} H\ell \left( 1-z_0, [(1-z_0) (1-\theta_-) +1-\theta_+] (2-\Delta) + \kappa_1 \kappa_2-\tilde{K}; \right.\\
&~~~~~~~~~~~~~~~~~~~~~~~~~~~\kappa_1+2-\Delta, \kappa_2+2-\Delta, 3-\Delta, 1-\theta_-; 1-z \Bigr),
\end{split}
\end{align}
for $z\sim 1 (r \sim \infty)$. Here $c_i$ and $d_i$ ($i=0,1$) are arbitrary constants and we define $\tilde K\equiv K+\kappa_1\kappa_2z_0$. For the angular equation (\ref{heun_angular}), its general solution is
\begin{align}
\begin{split}
S &= \tilde{c}_0 H \ell \left( u_0, \tilde{Q}; q_1, q_2, 1+m_1, \Delta-1; u \right)\\
&+\tilde{d}_0 z^{m_1} H\ell \left( u_0, \tilde{Q} -m_1 [u_0 (\Delta-1) +m_2+1];\right.\\ 
&~~~~~~~~~~~~~~~~~~~~~q_1-m_1, q_2 -m_1, 1-m_1, \Delta-1; u \Bigr)
\end{split}
\end{align}
around $u \sim 0$ ($\theta \sim 0$), and
\begin{align}
\begin{split}
S &= \tilde{c}_1 H\ell \left( \frac{u_0}{u_0-1}, \frac{-Q}{u_0-1}; q_1,q_2,1+m_2, \Delta-1; \frac{u_0-u}{u_0-1} \right)\\
&+\tilde{d}_1 \left( \frac{u_0-u}{u_0-1} \right)^{-m_2} H\ell \left( \frac{u_0}{u_0-1}, \frac{-Q-m_2 [u_0 (\Delta +m_1) -(1+m_1)]}{u_0-1}; \right.\\
&~~~~~~~~~~~~~~~~~~~~~~~~~~~~~~~~~~~~~q_1-m_2, q_2-m_2,1-m_2,\Delta-1; \frac{u_0-u}{u_0-1} \biggr)
\end{split}
\end{align}
around $u \sim u_0$ ($\theta \sim \pi/2$), where $\tilde{c}_i$ and $\tilde{d}_i$ ($i=0,1$) are arbitrary constants and $\tilde Q\equiv Q+q_1q_2u_0$. To ensure the ingoing and Dirichlet boundary conditions at $r=r_+$ and $r= \infty$, respectively, one has to impose the following boundary condition for $\Pi (z)$
\begin{align}
\Pi (z) \sim
\begin{cases}
(z-z_0)^{-\theta_+/2} &\text{for} \ z\to z_0 \ (r \to r_+),\\
(z-1)^{\Delta/2} &\text{for} \ z\to 1 \ (r \to \infty),
\end{cases}
\label{BC_PI}
\end{align}
and for $\Theta (u)$, the regular condition at $u=0$ ($\theta = 0$) and $u = u_0$ ($\theta = \pi/2$) is
\begin{align}
\Theta (u) \sim 
\begin{cases}
u^{|m_1|/2} &\text{for} \ u\to 0,\\
(u-u_0)^{|m_2|/2} &\text{for} \ u\to u_0.
\end{cases}
\label{BC_TH}
\end{align}
To satisfy the boundary condition (\ref{BC_PI}), $R(z)$ should take the following form at the boundaries
\begin{align}
R &\sim R_{\rm in} (\omega, z) \ \text{for} \ z \sim z_0,
\label{BH_BC}\\
R &\sim R_{\rm AdS} (\omega,z) \ \text{for} \ z \sim 1.
\end{align}
For $u \sim 0$, on the other hand, $S(u)$ takes the following form
\begin{align}
S =
\begin{cases}
&H\ell (u_0, \tilde Q; q_1, q_2, 1+m_1, \Delta-1; u)
~~~~~~~~~~~~~~~~~~~~~~~~~\text{for} \ m_1 \geq 0,\\
&z^{-m_1} H\ell \Bigl( u_0, -m_1 [u_0 (\Delta-1) + 1+m_2] + \tilde Q;\\
&~~~~~~~~~~~~~~~q_1-m_1, q_2-m_1, 1-m_1, \Delta-1; u \Bigr)
~~~~~~~ \text{for} \ m_1 \leq 0,
\end{cases}
\end{align}
and for $u \sim u_0$, $S(u)$ is
\begin{align}
S =
\begin{cases}
&\displaystyle
H\ell \left(\frac{u_0}{u_0-1}, \frac{q_1 q_2 u_0 -\tilde Q}{u_0-1}; q_1, q_2, 1+m_2, \Delta-1; \frac{u_0-u}{u_0-1} \right)
~~~\text{for} \ m_2 \geq 0\\
&\displaystyle
\left( \frac{u_0-u}{u_0-1} \right)^{-m_2} H\ell \biggl( \frac{u_0}{u_0-1}, \frac{-m_2 [u_0 (m_1+\Delta) -1-m_1]}{u_0-1} + \frac{q_1 q_2 u_0 -\tilde Q}{u_0-1};\\
&\displaystyle
~~~~~~~~~~~~~~~~~q_1-m_2, q_2-m_2, 1-m_2, \Delta-1; \frac{u_0-u}{u_0-1} \biggr)
~~~~~\text{for} \ m_2 \leq 0.
\end{cases}
\label{S_pi_BC}
\end{align}
In the following sections, we use the {\it Wolfram Mathematica} to search for the eigenvalues $\lambda_{lm_1m_2n}$ and QN frequencies $\omega_{lm_1m_2n}$ for which the boundary conditions (\ref{BH_BC}) - (\ref{S_pi_BC}) are satisfied.
\begin{figure}[h]
\centering
\includegraphics[width=12cm]{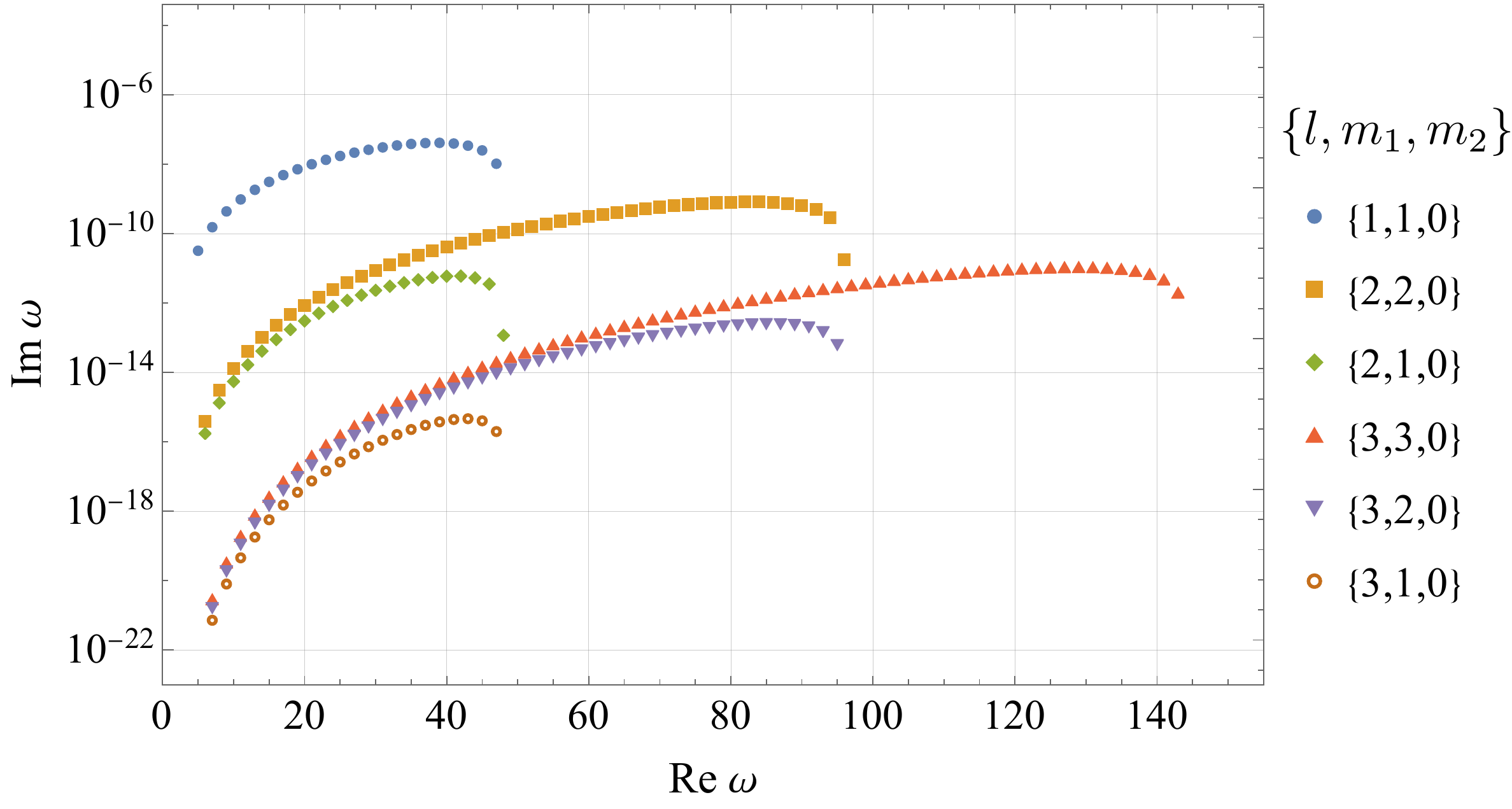}
\caption{Plot of QN frequencies for $l=1,2,3$ modes. We set $a_1=a_2=4\times 10^{-4}$, $\mathcal M =10^{-5}$ and $\mu =10^{-2}$.}
\label{ell}
\end{figure}
\begin{figure}[h]
\includegraphics[width=7.5cm]{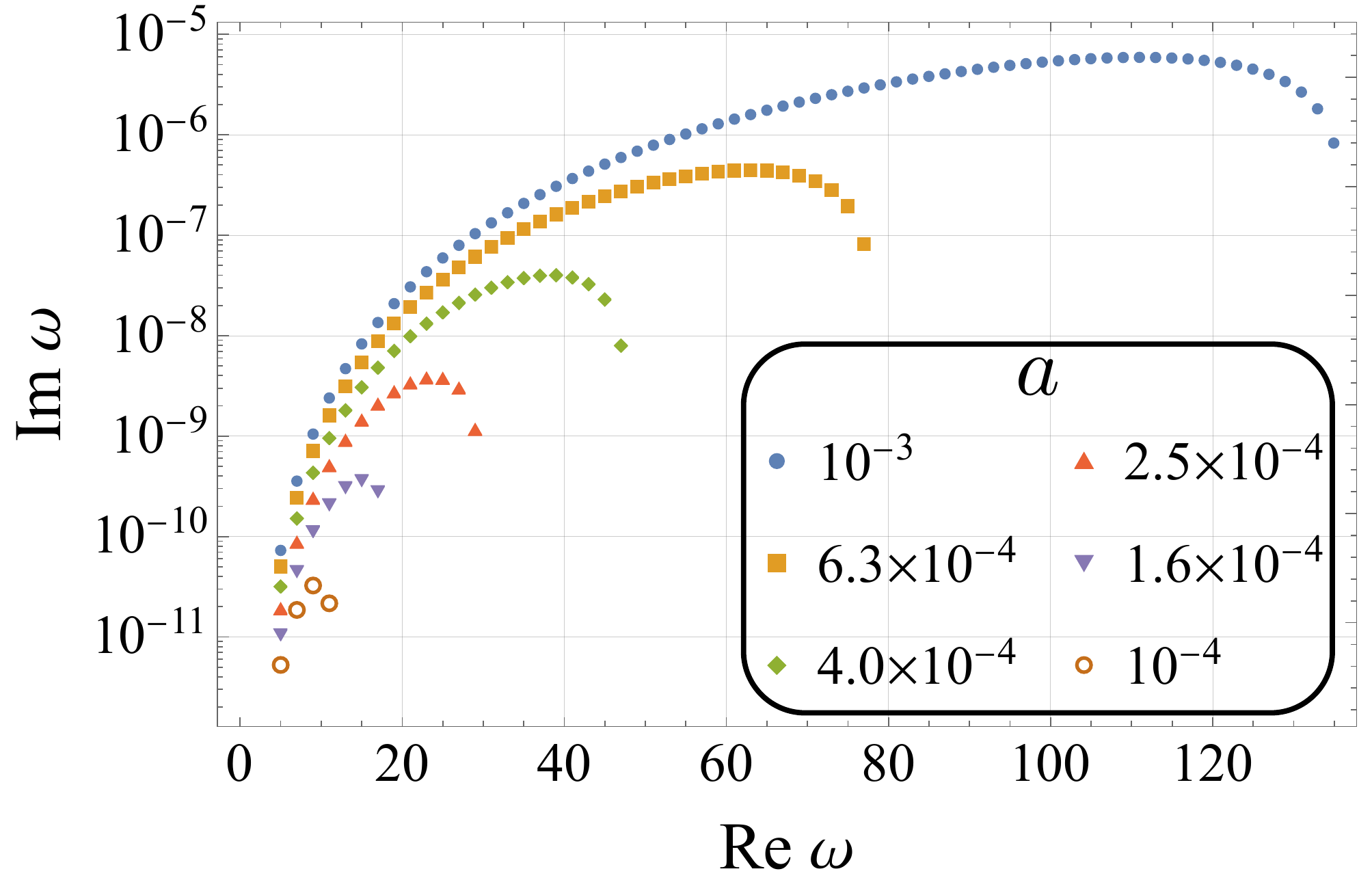}
\includegraphics[width=8cm]{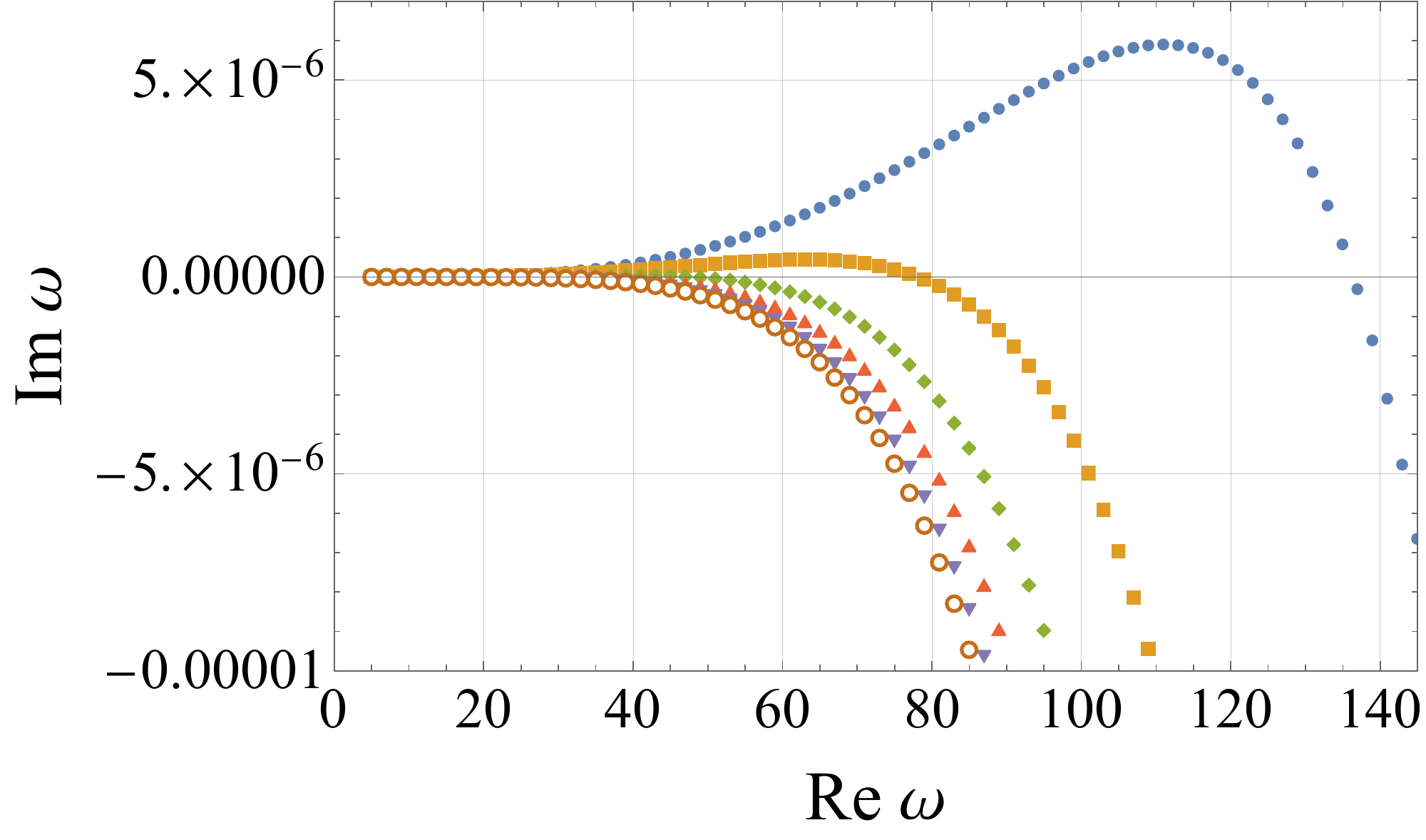}
\caption{Plot of QN frequencies for various spin parameters. In the left and right panels, the imaginary part is shown in log and linear scale, respectively. Each marker indicates the complex value of each QN frequency. We set $\mathcal M =10^{-5}$ and $\mu =10^{-2}$, and the spin parameters are in the range of $10^{-4} \leq a(=a_1=a_2) \leq 10^{-3}$. The angular modes are fixed as $(l,m_1,m_2) = (1,1,0)$ that is equivalent to $(1,0,1)$.}
\label{QNM_eq}
\end{figure}
\section{Stability analysis for equal spins ($a_1=a_2$)}
\label{sec:equal}
In this section, we numerically investigate the QN frequencies for a \kdsf black hole with $a_1=a_2 \equiv a$. For equal spins, the angular equation reduces to the hypergeometric differential equation, and one can obtain the analytic expression of the eigenvalue $\lambda$ \cite{Aliev:2008yk}
\begin{equation}
\lambda = (1-a^2) \left[ l (l+2) -2 \omega a (m_1+m_2) -a^2 (m_1+m_2)^2 \right] + a^2 \omega^2 +a^2 \Delta (\Delta-4),
\label{separation_constant}
\end{equation}
where $l = 0, 1, 2,...$ is the angular mode. 
We can obtain QN frequencies by solving the radial equation (\ref{heun_radial}) with $\lambda = \lambda (\omega)$ given in (\ref{separation_constant}) and by searching for $\omega = \omega_{lm_1m_2n}$ at which the obtained solution satisfies the boundary condition (\ref{BC_PI}).

\subsection{Small black holes ${\cal M} \ll 1$}
\label{sec:equal_small_BH}
For small black holes, QN frequencies are localized near the real axis of the complex frequency plane, $|\text{Re} (\omega_{lm_1m_2n})| \gg |\text{Im} (\omega_{lm_1m_2n})|$, due to trapped modes in the AdS boundary. Also, the superradiant instability is caused by the resonance between the ergoregion and AdS boundary. The stability of the black hole can be read from the sign of the imaginary part of QN frequencies, and $\text{Im} (\omega_{lm_1m_2n}) > 0$ means that the background spacetime is unstable against linear perturbations.
For small \kdsf black holes, the unstable QN modes satisfy the following condition
\begin{equation}
\label{superradiant_cond}
\text{Re}({\omega_{lm_1m_2n}}) < m_1 \Omega_{+,1} +m_2 \Omega_{+,2} \equiv \Omega.
\end{equation}
Here it is natural to ask which overtone leads to the most significant instability when multiple overtones satisfy the above condition.
\begin{figure}[h]
\centering
\includegraphics[width=12.5cm]{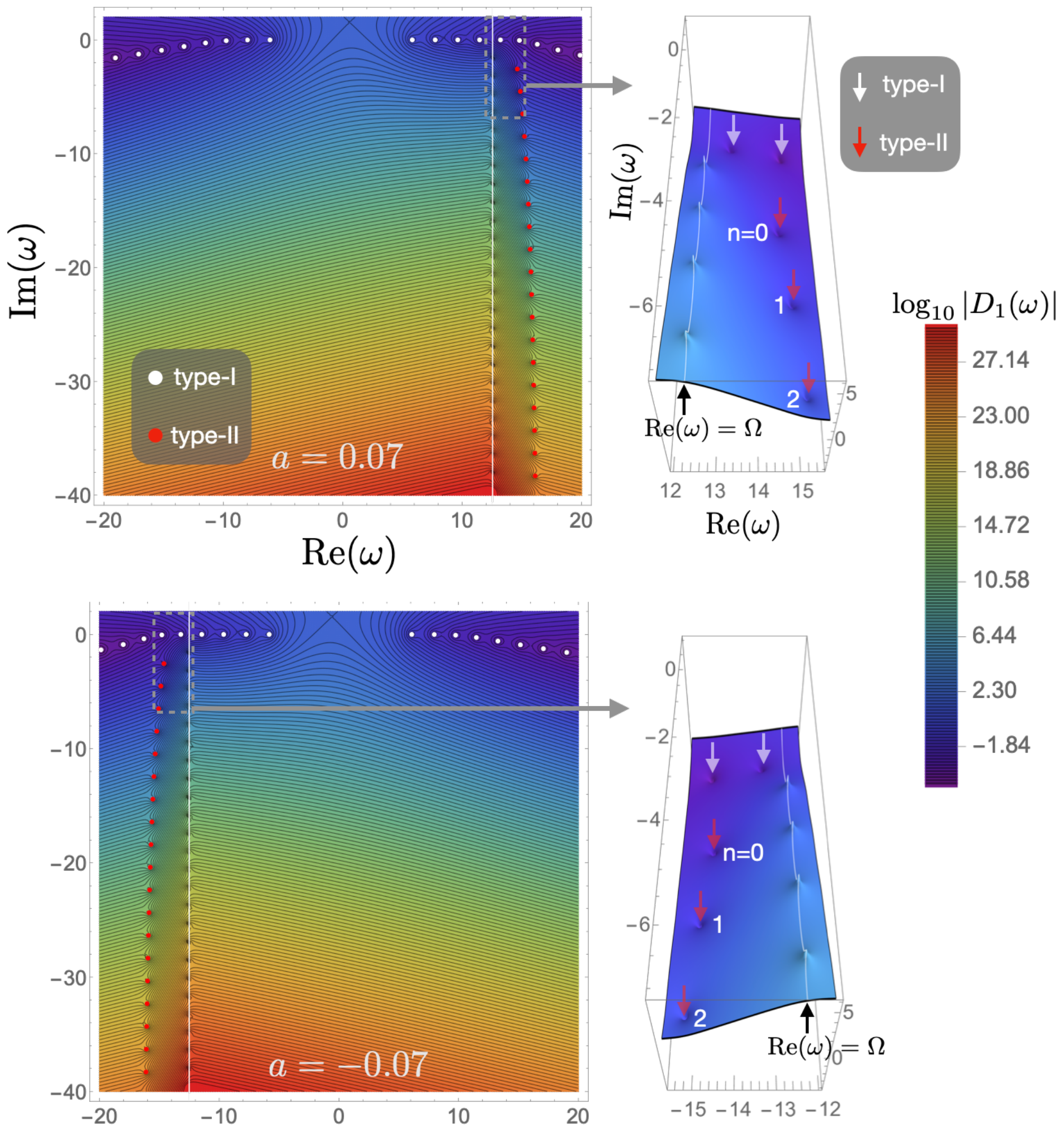}
  \caption{Plots of the coefficient $D_1 (\omega)$ defined in (\ref{define_d1}). The parameters are set to $a= \pm 0.07$, $M = 0.01$, and $(l,m_1,m_2) = (2,1,1)$. The type-II modes are numbered by non-negative integers, $n$, in ascending order of the damping rates.}
\label{global_small_equal}
\end{figure}
To see this, we first investigate QN modes for $l=1,2,3$ in Figure \ref{ell}, and it is found that $l=1$ mode leads to the most significant instability.
We then study the spin-dependence of QN modes with $(l,m_1,m_2) = (1,1,0)$, and the result is shown in Figure \ref{QNM_eq}. 
It is shown that rapid rotations destabilize the \kdsf black holes, and the peak of $\text{Im} (\omega_{lm_1m_2n})$ is slightly below the superradiant frequency of $\Omega$.
We also confirm that our result shown in Figure \ref{QNM_eq} is consistent with the condition of the superradiant instability (\ref{superradiant_cond}) as
$\text{Re}({\omega_{lm_1m_2n}})<136.422$ and $\text{Re}({\omega_{lm_1m_2n}})<11.795$ for $a=10^{-3}$ and $a=10^{-4}$, respectively. In the following, we will omit the subscript $+$ from $\Omega_{+,i}$.
Besides the QN modes localized near the real axis of $\omega$, highly damped QN modes are localized near the line of $\text{Re} (\omega) = \Omega$ in the complex frequency plane, as is shown in Figure \ref{global_small_equal}. The coefficient $D_1(\omega)$, defined as
\begin{equation}
R_{\rm BH} (\omega,z) = C_1(\omega) R_{\rm AdS} (\omega, z) + D_1(\omega) R_{\rm Div} (\omega, z),
\label{define_d1}
\end{equation}
is plotted in Figure \ref{global_small_equal}, and it vanishes at $\omega = \omega_{lm_1m_2n}$.
Hereinafter, we call QN modes localized near the real axis of $\omega$ and those localized near the superradiant frequency {\it type I} and {\it type II}, respectively.
\begin{figure}[h]
\centering
\includegraphics[width=15cm]{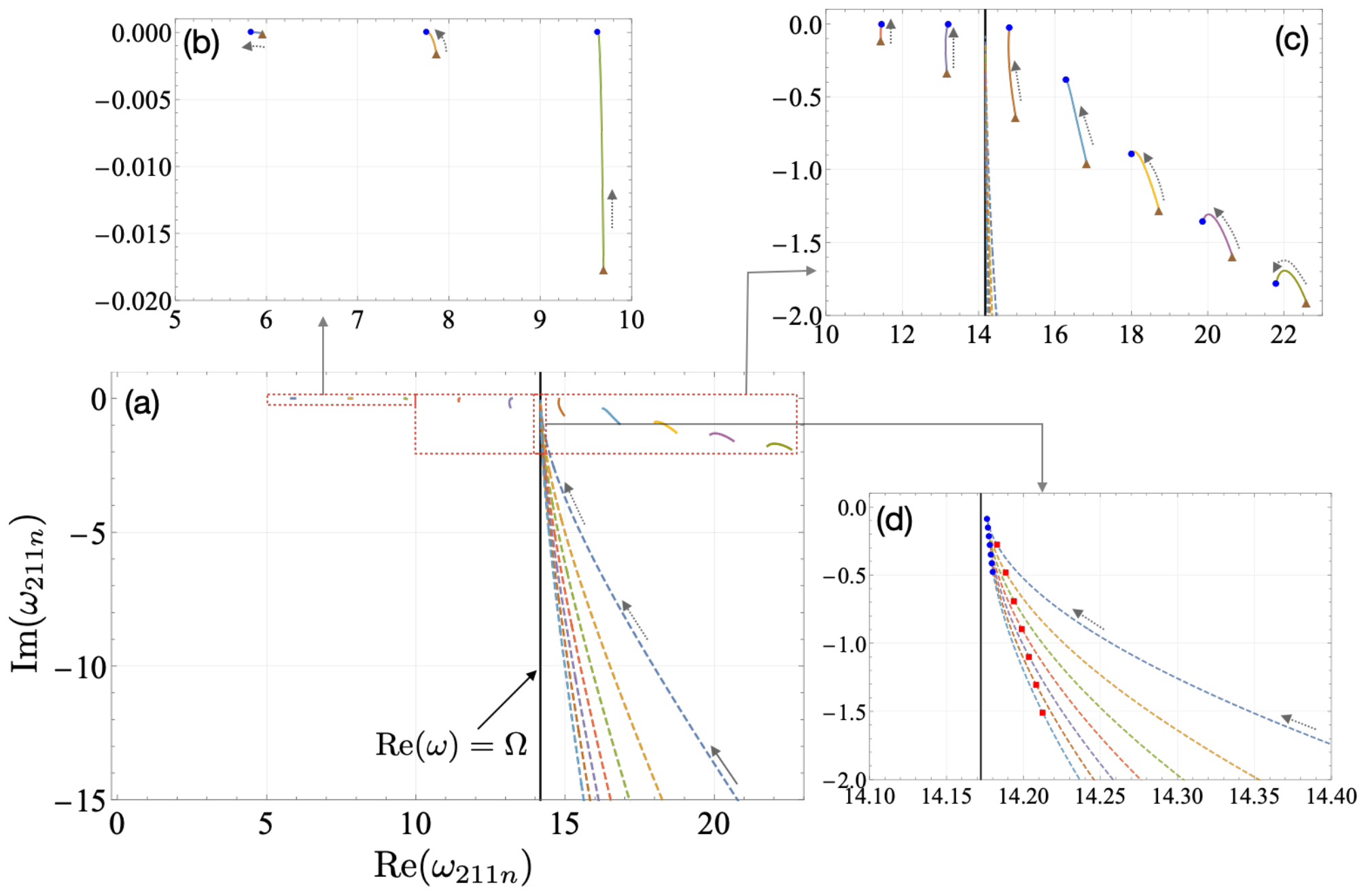}
  \caption{Trajectories of QN frequencies for $M=0.01$, $(l,m_1,m_2) = (2,1,1)$, and $\mu = 0.01$. The solid and dashed lines indicate the type-I and type-II QN modes, respectively. The spin parameter $a$ runs from $a=10^{-4}$ (brown triangles) to $a\simeq 0.999996 a_{\rm max}$ (blue dots) in (b) and (c). The spin parameter runs up to $a = 0.99999 a_{\rm max}$ (blue dots) in (d). The maximum spin parameter is $a_{\rm max} = 0.070536282...$ and the red squares in (d) indicate the QN frequencies at $a=0.07$. The arrows indicate the direction in which the spin parameter, $a$, increases.}
\label{flow_QNM_light}
\end{figure}
The type-I modes are caused by the resonance between the AdS barrier and the angular momentum barrier. On the other hand, type-II modes may be relevant to the thermality of the \kdsf black hole as their separation in frequency space is nearly equal to the surface gravity of \kdsf black hole $2 \pi T_{\rm H}$, and the real part of the type-II modes can be interpreted as the chemical potential. Figure \ref{flow_QNM_light} shows the trajectories of the type-I and type-II modes with respect to the change of the Hawking temperature. One can see that the lower the Hawking temperature of the \kdsf black hole is, the stronger the localization of type-II modes at the superradiant frequency is (see Figure \ref{small_real_imag}). Also, the separation of the imaginary part of QN frequencies, defined as $\Delta \text{Im} (\omega_{l m_1 m_2 n}) \equiv \text{Im} (\omega_{l m_1 m_2 n} - \omega_{l m_1 m_2 (n+1)})$, approaches $2 \pi T_{\rm H}$ in the zero-temperature limit. It looks like the behaviors of highly damped QN modes are insensitive to $\mu$ at least for $0.01 \leq \mu \leq 10$.

In the next subsection, we will see the massive case, ${\cal M} \gtrsim 1$, for which the lowest Hawking temperature is nonzero and finite due to the upper bound on the two spins ($a_i \leq 1$ with $i=1,2$). We will show the structure of the type-II modes in the complex frequency plane for the lowest Hawking temperature.

\begin{figure}[h]
\centering
\includegraphics[width=14cm]{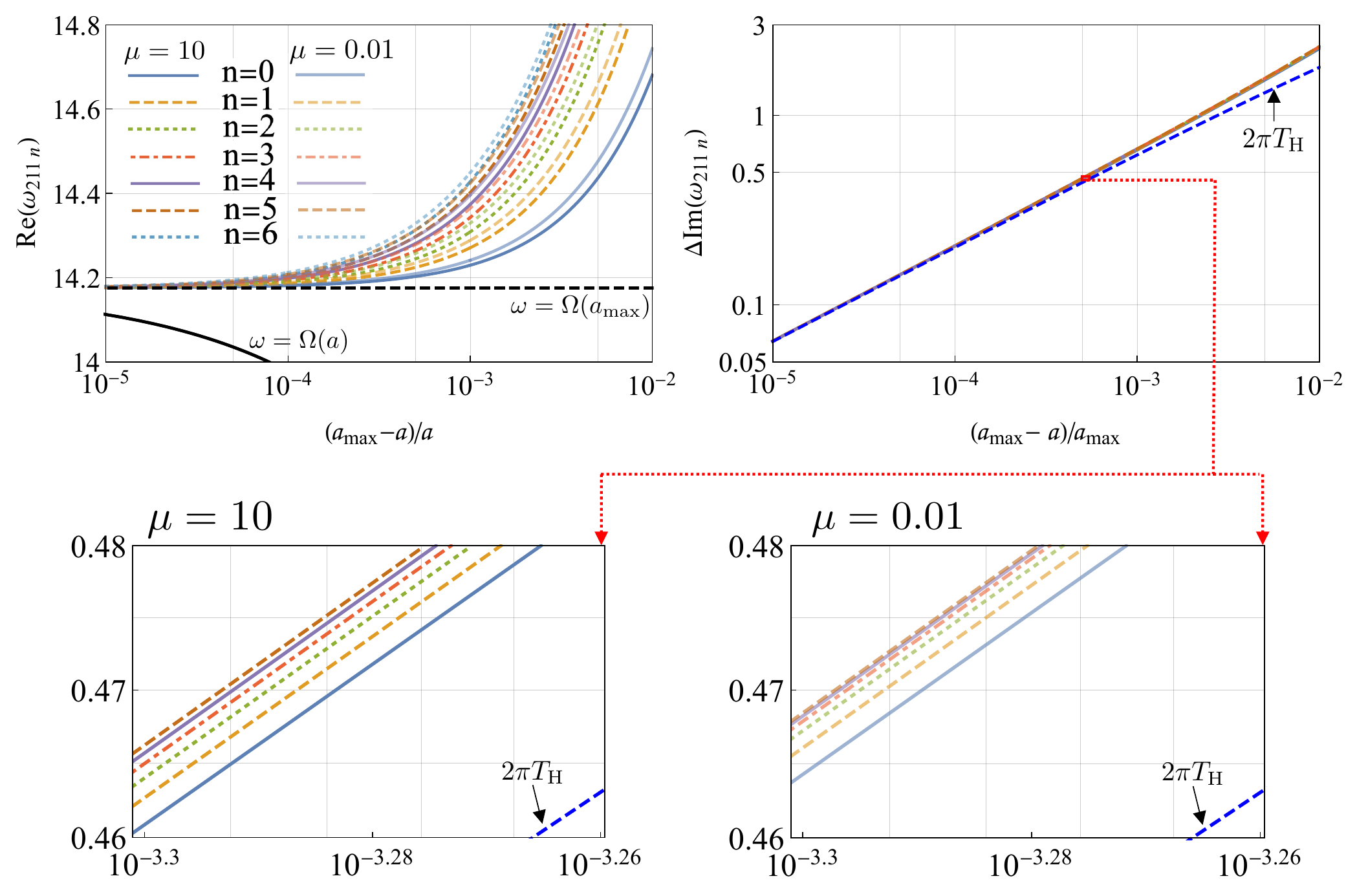}
\caption{
The real part and the separation in the imaginary part of QN frequencies are shown with respect to $(a_{\rm max} - a)/a_{\max}$ in the upper left and upper right panels, respectively. 
We change the opacity of each line to distinguish the case of $\mu=0.01$ (transparent lines) and $\mu=10$ (opaque lines). The other parameters are set to $M=0.01$ and $(l,m_1,m_2) = (2,1,1)$.
In the left panel, the black solid line indicates the superradiant frequency, $\Omega$, that depends on $a$, and the black dashed line shows the value of $\Omega$ at the extremal case ($a=a_{\rm max}$). The blue dashed line in the right panel is $2 \pi T_{\rm H} (a)$. As a reference, the region of $10^{-3.3} \leq (a_{\rm max} -a)/a_{\rm max} \leq 10^{-3.26}$ is zoomed in and displayed in the lower panels.}
\label{small_real_imag}
\end{figure}

\subsection{Large black holes ${\cal M} \gtrsim 1$}
\label{sec:equal_large_BH}
\begin{figure}[b]
\centering
\includegraphics[width=12cm]{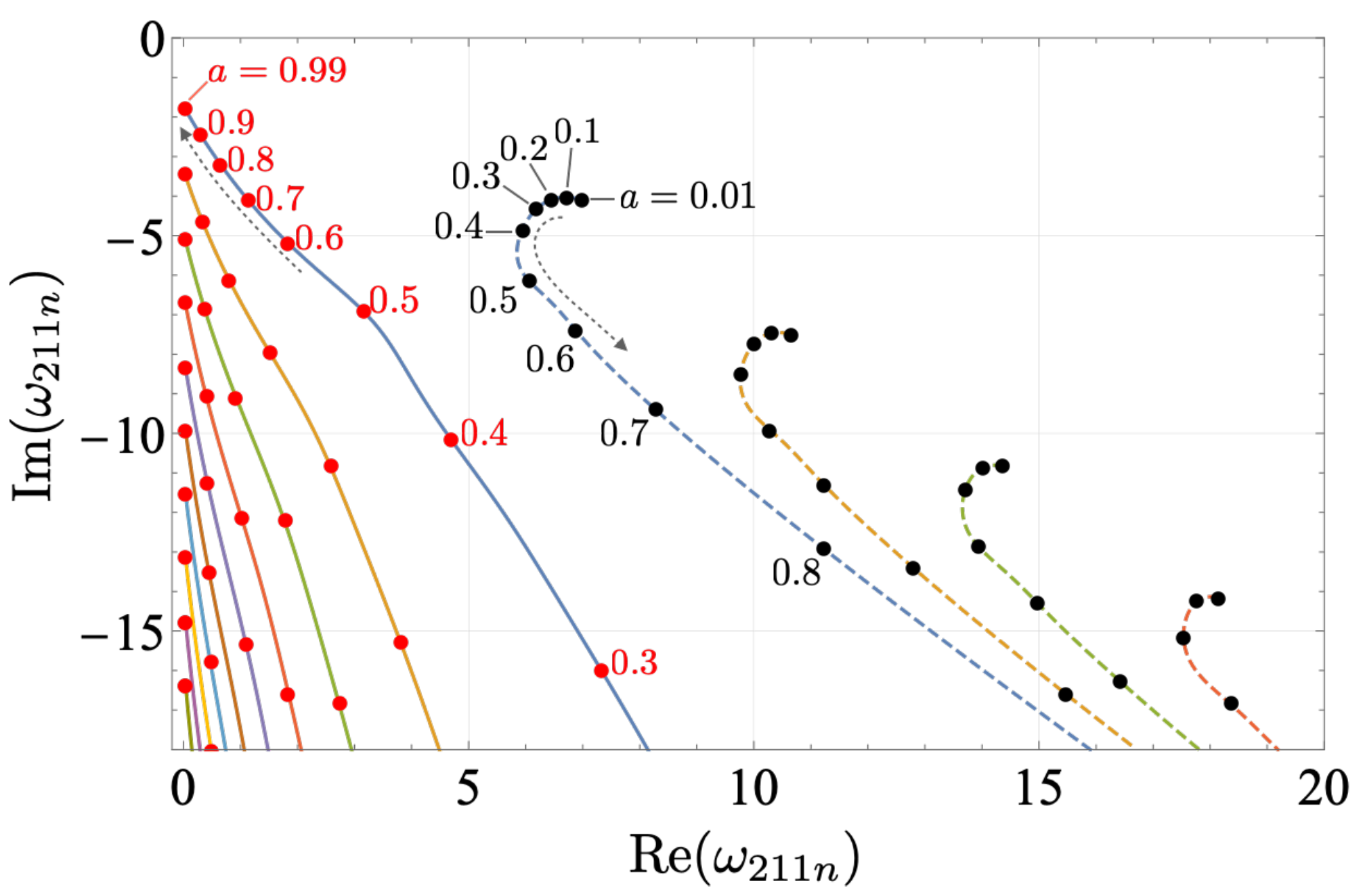}
  \caption{Trajectories of QN frequencies for $M=5$ and $(l,m_1,m_2) = (2,1,1)$. The spin parameter $a$ runs from $a=0.01$ to $a=0.99$. The dashed and solid lines indicate the type-I and type-II QN modes, respectively. The arrows indicate the direction in which the spin parameter, $a$, increases.}
\label{flow_QNM}
\end{figure}

\begin{figure}[h]
\centering
\includegraphics[width=15cm]{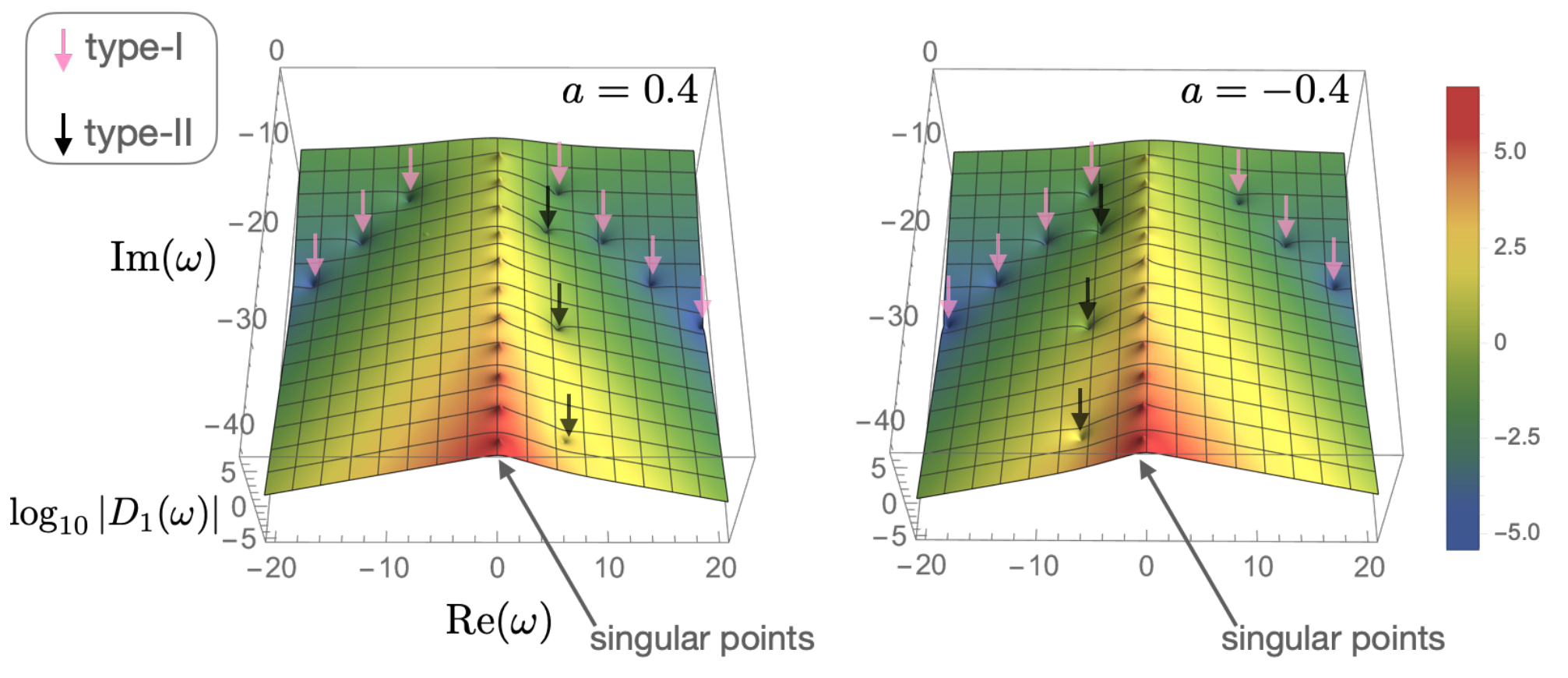}
\caption{3D plots of the coefficient $D_1 (\omega)$ in the complex frequency plane. The zeros of $D_1$ correspond to the complex QN frequencies, $\omega = \omega_{lm_1m_2n}$, and the pink and black arrows indicate the type-I and type-II modes, respectively.
Here we set $(l,m_1,m_2) = (2,1,1)$, $M=5$, and $\mu = 0.01$. The spins are set to $a= 0.4$ (left) and $a=-0.4$ (right).}
\label{FIG3d}
\end{figure}

Massive Kerr-AdS${}_5$ black holes (${\cal M} \gtrsim 1$) are stable against linear perturbations, which is equivalent to $\text{Im} (\omega_{lm_1m_2n})<0$ for all modes, since the cavity between the angular momentum barrier of the black hole and the AdS boundary to cause the resonant instability does not exist for massive black holes. The trajectories of QN modes for $M=5$ is shown in Figure \ref{flow_QNM}. Note that type-I modes (dashed lines) correspond to the QN modes that localize near the real axis of $\omega$ in the small mass regime (c.f. Sec. \ref{sec:equal_small_BH}).
In that case, the type-I modes are caused by the resonance in the AdS boundary, and thus, the type-I modes periodically appear near the real axis of $\omega$, which is similar to the normal modes of a vibrating string. In the massive case, on the other hand, type-I modes are highly suppressed especially for $a\to 1$, while type-II modes localize at $\text{Re} (\omega) = \Omega$ (see Figure \ref{flow_QNM}). Note that in the limit of $a \to 1$, the superradiant frequency also vanishes $\Omega \propto (1-a^2) \to 0$. We also find that the type-II modes appear from the region of $\text{Re} (\omega) > \Omega$ for $a> 0$ and they do from the region of $\text{Re} (\omega) < \Omega$ for the counter rotations ($a< 0$), as is shown in Figure \ref{FIG3d}, where the absolute value of $|D_1(\omega)|$ is shown in the log-scale.
\begin{figure}[h]
\centering
\includegraphics[width=14.5cm]{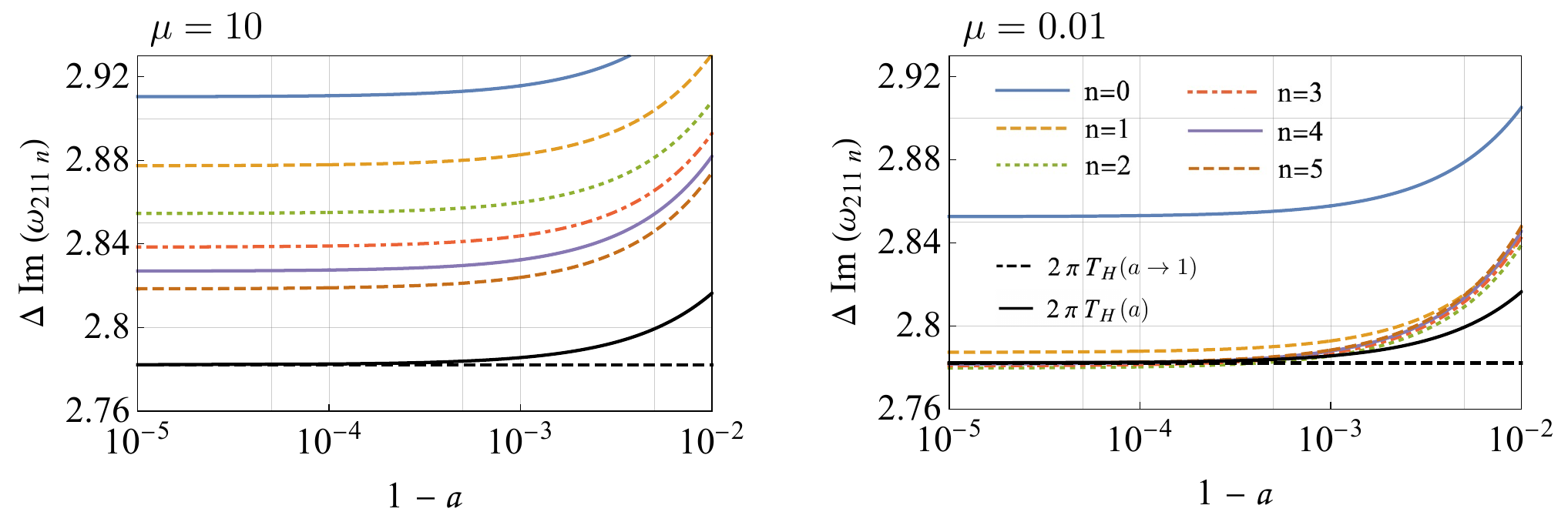}
\caption{Plots of $\Delta \text{Im} (\omega_{211n})$ for $\mu =10$ (left)
and $\mu = 0.01$ (right). The black line indicates 2$\pi T_{\rm H} (a)$, and black dashed line indicates 2$\pi T_{\rm H} (a\rightarrow 1)$. The other parameters are set to $M=10$ and $(l,m_1,m_2)=(2,1,1)$.}
\label{FIGa}
\end{figure}

\begin{figure}[t]
\centering
\includegraphics[width=14cm]{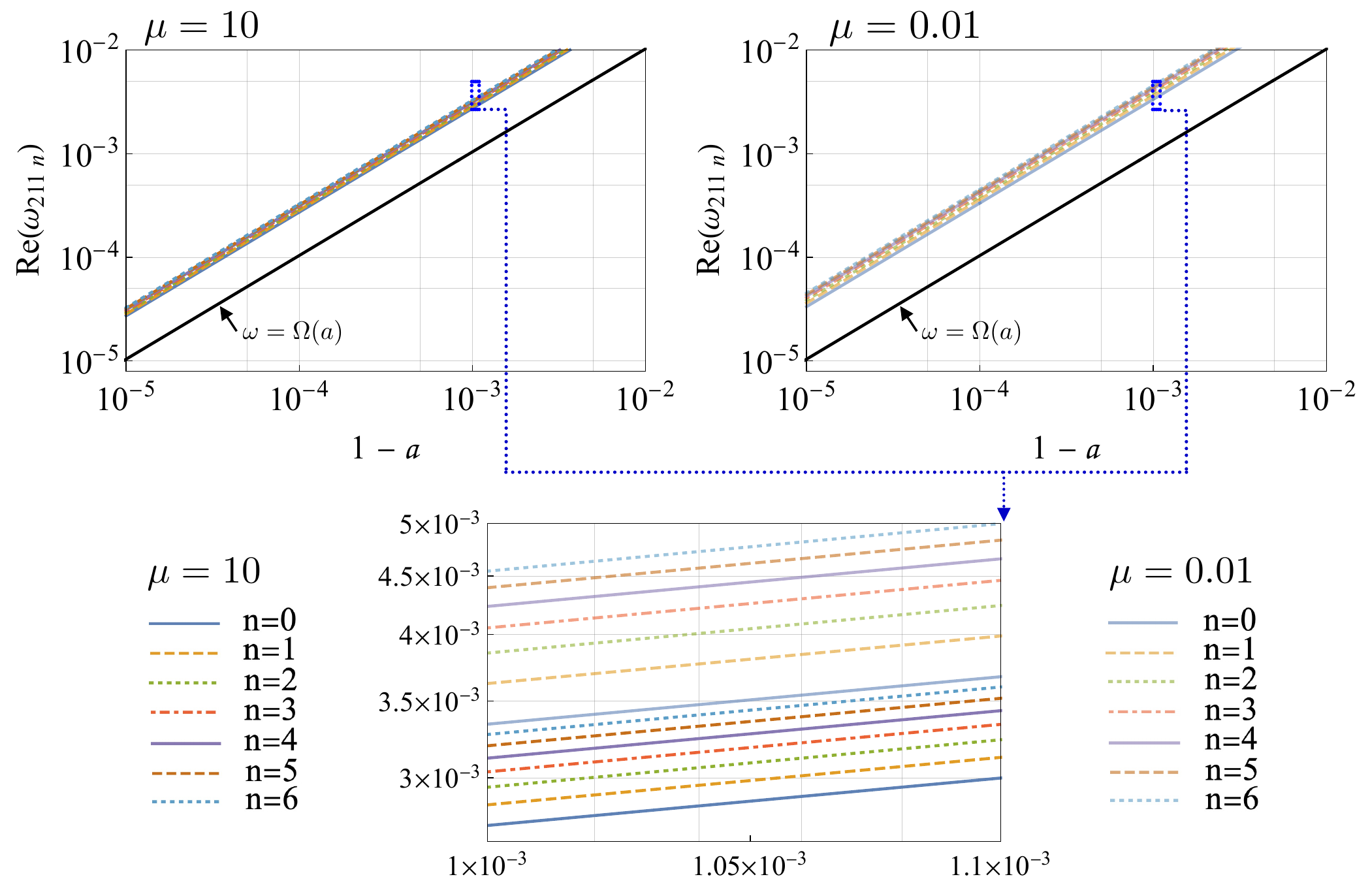}
\caption{Plots of $\text{Re} (\omega_{211n})$ for $\mu =10$ (upper left)
and $\mu = 0.01$ (upper right). The other parameters are set to $M = 10$ and $(l,m_1,m_2) =
(2,1,1)$. We change the opacity of each line to distinguish the case of $\mu= 0.01$ (transparent lines)
and $\mu=10$ (opaque lines). The region of $1.0\times10^{-3}\leq1-a\leq 1.1\times 10^{-3}$ is zoomed in and displayed in the lower panel for comparison.}
\label{FIGa39}
\end{figure}

We confirm that the separation $\Delta \text{Im} (\omega_{lm_1m_2n})$ approaches $2 \pi T_{\rm H}$ for $a \to 1$ as is shown in Figure \ref{FIGa}. It looks like the separation converges to $2 \pi T_{\rm H}$ for highly damped modes (higher overtone number $n$) though it is dispersive, and its convergence is weaker when the mass of the scalar field, $\mu$, is massive. This implies that the convergence of $\Delta \text{Im} (\omega_{lm_1m_2n})$ to $2 \pi T_{\rm H}$ for $n \gg 1$ is delayed or that the thermality of the \kdsf black hole would be disturbed by the mass of surrounding fields. Figure \ref{FIGa39} shows that nevertheless the real part of type-II modes strongly converges to the superradiant frequency $\Omega$ in the ultra-spinning limit for both small and large $\mu$.
As a final remark of this section, we have numerically computed the type-II QN frequencies up to finite overtone numbers. There is still a possibility that even high-temperature \kdsf black holes have the thermal structure of their QN modes at $n \to \infty$, i.e., the real part and the separation of the imaginary part of the type-II modes match $\Omega$ and $2 \pi T_{\rm H}$, respectively, in the highly damped limit.

\section{Stability analysis for unequal spins ($a_1> a_2$)}
\label{sec:unequal}
In this section, we numerically investigate the configuration of QN modes in the complex frequency plane for \kdsf black holes with unequal-spin parameters $a_1 > a_2$. In this case, the symmetry of spacetime reduces to $U(1) \times U(1)$ while it is enhanced to $U(2)$ for $a_1=a_2$. We investigate how the reduction of the symmetry affects the superradiant instability in Sec. \ref{sec:unequal_small} and study the structure of highly damped QN modes for unequal spins in Sec. \ref{sec:unequal_large}.

\subsection{Small black holes ${\cal M} \ll 1$}
\label{sec:unequal_small}
\begin{figure}[t]
\centering
\includegraphics[width=15cm]{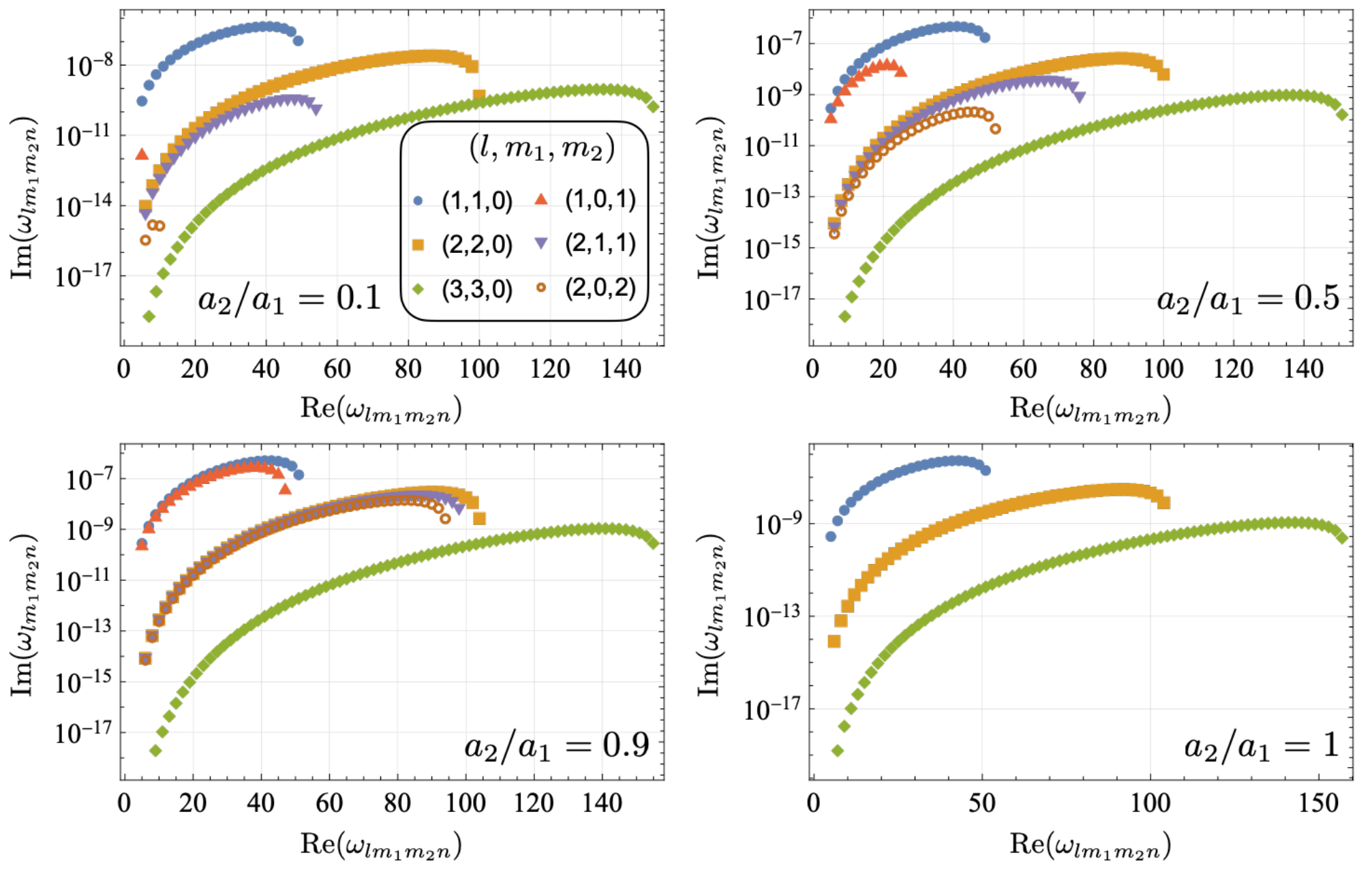}
  \caption{QN frequencies of $l=1,2,3$ modes for various ratio of spin parameters ($a_2/a_1= 0.1,$ $0.5$, $0.9$, and $1$). We set the background metric as $M =10^{-5}$, $\mu =10^{-2}$, and $a_1=10^{-4}$.}
    \label{ell_neq}
  \end{figure}

\begin{figure}[htbp]
\centering
\includegraphics[width=10cm]{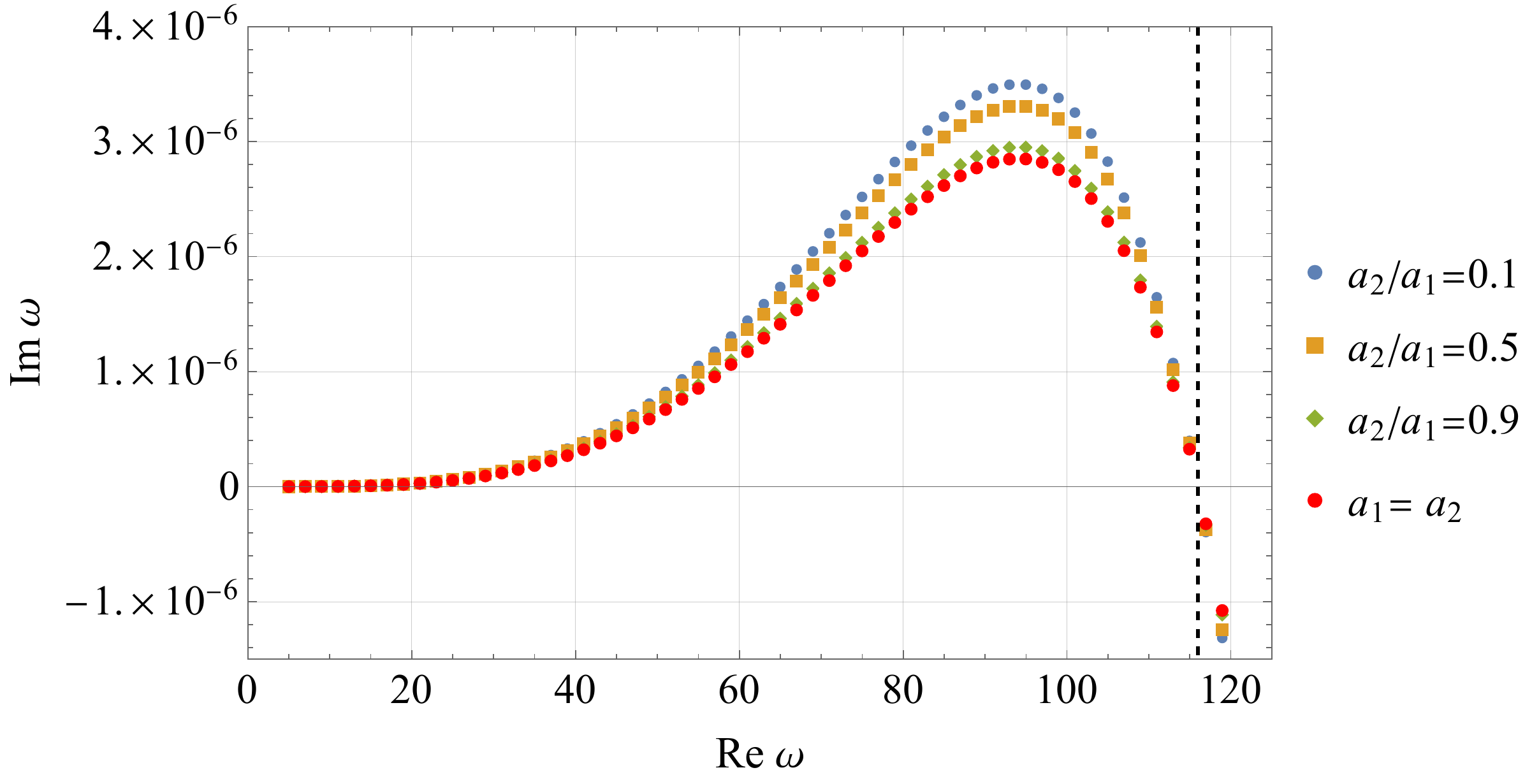}
  \caption{Plot of QN frequencies for various values of the spin ratio. Each marker indicates the complex value of each QN frequency. We set $\mathcal M =10^{-5}$, $\mu =10^{-2}$, and superradiant frequency $\Omega=116$, and the spin ratio is set to $a_2/a_1 = 1$, $0.9$, $0.5$, and $0.1$. The angular modes are fixed as $(l,m_1,m_2) = (1,1,0)$. The black dashed line indicates the superradiant frequency.}
    \label{QNM_neq}
  \end{figure}

To see which harmonic mode is the most significant to destabilize the system, we numerically compute the QN frequencies for various values of the spin ratio, $a_2/a_1$. We then numerically confirm that the most dominant instability is caused by the mode of $(l,m_1,m_2) = (1,1,0)$ for $a_1 > a_2$ as is shown in Figure \ref{ell_neq}. Therefore, we consider the instability of $(1,1,0)$ only and investigate how the reduction of the symmetry of spacetime affects the instability\footnote{The opposite hierarchy between $a_1$ and $a_2$ $(a_2 / a_1> 1)$ results in the suppression of the instability of $(1,1,0)$, and then the instability caused by $(1,0,1)$ mode becomes dominant.}. We compute the QN frequencies by fixing the ADM mass, the mass of the scalar field, and the superradiant frequency in Figure \ref{QNM_neq}.
As a result, we find that the instability is more significant when the spin ratio $a_2/a_1$ is smaller. 
Our result implies that the symmetry reduction of the \kdsf black hole leads to the enhancement of the superradiant instability when ${\cal M}$, $\mu$, and $\Omega_1$ are fixed.

\begin{figure}[t]
\centering
\includegraphics[width=14cm]{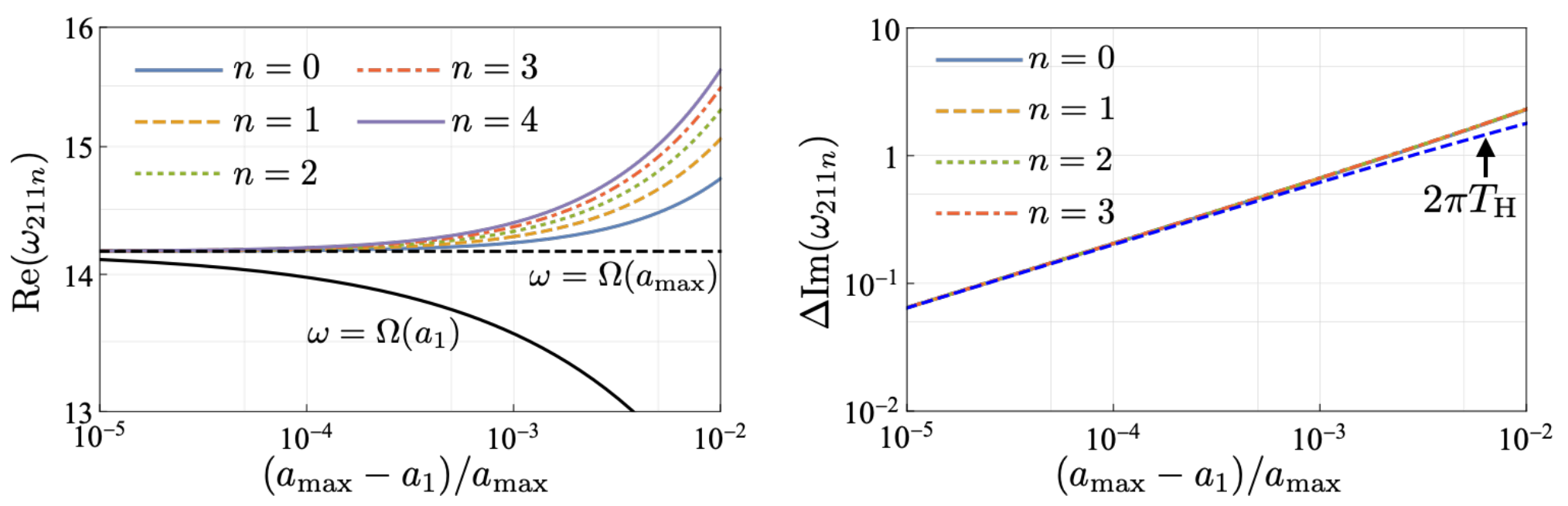}
  \caption{The real part and the separation of the imaginary part of QN frequencies are shown with respect to $(a_{\rm max} - a_1)/a_{\max}$ in the left and right panels, respectively. We set $a_2/a_1 =0.9$, $M=0.01$, and $\mu = 0.01$. In the left panel, the black solid line indicates the superradiant frequency, $\Omega$, that depends on $a_1$, and the black dashed line shows the value of $\Omega$ at the extremal case ($a_1=a_{\rm max}$). The blue dashed line in the right panel shows $2 \pi T_{\rm H}(a_1)$.}
\label{small_real_imag_neq}
\end{figure}

We found that the structure of type-II modes exhibits the thermodynamic nature of the \kdsf black hole in the low-temperature or ultra-spinning limits with $a_1=a_2$ (see Sec. \ref{sec:equal_small_BH} and \ref{sec:equal_large_BH}). We here investigate if the thermal interpretation of the type-II modes holds even for unequal spins ($a_1 \neq a_2$). 
The left panel in Figure \ref{small_real_imag_neq} shows that the real part of type-II modes approaches the superradiant frequency for lower $T_{\rm H}$, which is similar to the case of small \kdsf black holes with equal spins (see Figure \ref{small_real_imag}). From the right panel in Figure \ref{small_real_imag_neq}, one can also see that $\Delta \text{Im} (\omega_{211n})$ matches $2 \pi T_{\rm H}$ in the low-temperature limit. From those results, we confirm that the thermodynamic nature of the \kdsf black hole we observed for the equal spin case still holds even for unequal spins.

\subsection{Large black holes ${\cal M} \gtrsim 1$}
\label{sec:unequal_large}
\begin{figure}[t]
\centering
\includegraphics[width=14.5cm]{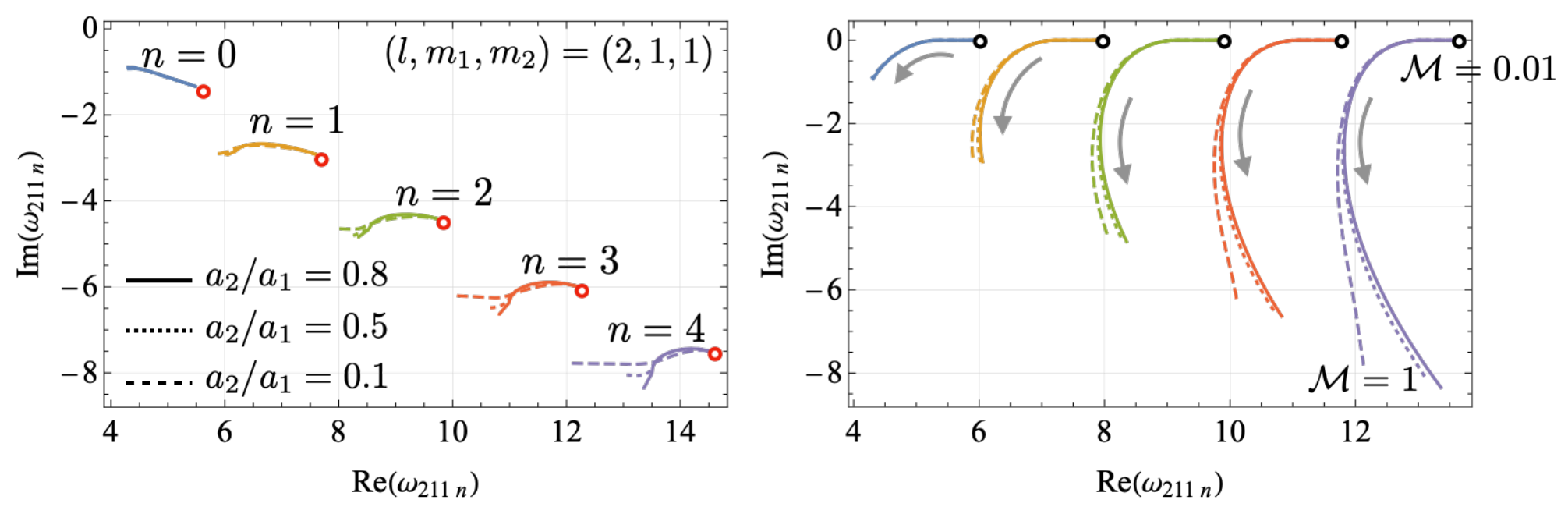}
 \caption{Trajectories of QN frequencies in the complex frequency plane. In the left panel, we set ${\cal M} = 1$ and the Hawking temperature changes from the lower value of $T_{\rm H} = 0.01$ to the maximum temperature while fixing the spin ratio $a_2/a_1$. The red open circles indicate the QN frequencies for which $T_{\rm H}$ takes the maximum value. In the right panel, the Hawking temperature is fixed at $T_{\rm H} = 0.01$ and the ADM mass changes from ${\cal M} = 0.01$ (black open circles) to ${\cal M} = 1$. The arrows indicate the direction in which the ADM mass increases. The harmonic mode is set to $(l,m_1,m_2) = (2,1,1)$.}
\label{unequal_stabilization}
\end{figure}

We here investigate the type-II QN frequencies of a large black hole ${\cal M} \gtrsim 1$ with unequal spins. The \kdsf spacetime is stable for ${\cal M} \gtrsim 1$ in the case of unequal spins, $a_1 > a_2$, as well. In Figure \ref{unequal_stabilization}, the QN frequencies are shown up to the 4th overtone. We can see that each QN frequency has a bent path when the Hawking temperature changes while the other quantities are fixed (left panel in Figure \ref{unequal_stabilization}). As a \kdsf black hole becomes massive, the background spacetime is getting stabilized, which can be seen in the right panel of Figure \ref{unequal_stabilization} and is similar to the equal spin case.
\begin{figure}[htbp]
\centering
\includegraphics[width=14.5cm]{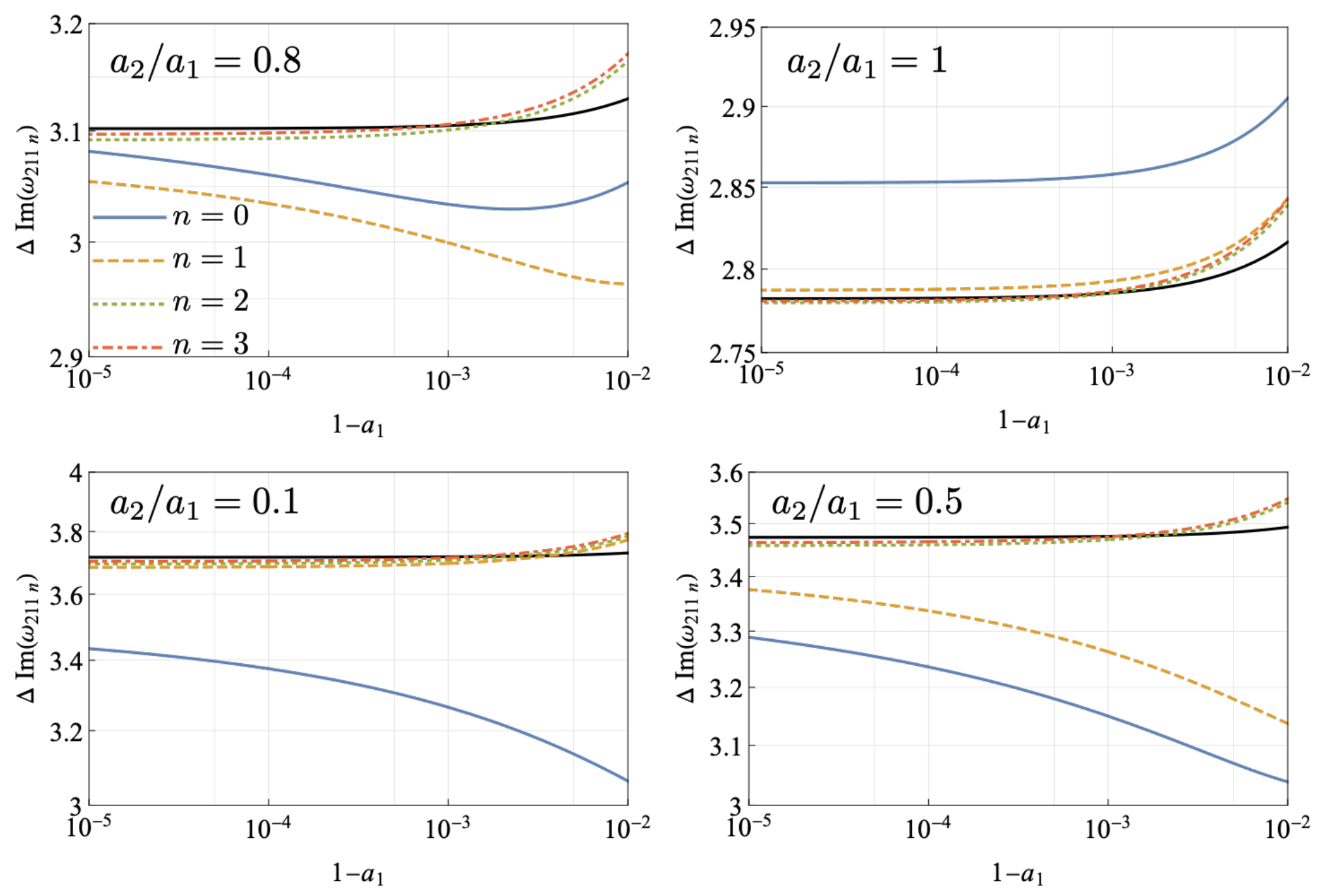}
\caption{The separation of the imaginary part of the type-II QN frequencies is shown as a function of $1-a_1$ in the ultra-spinning limit. The harmonic mode is set to $(l,m_1,m_2) = (2,1,1)$ and $M=10$. The black solid lines show the values of $2 \pi T_{\rm H}(a_1)$.}
\label{unequal_Hawking_temp}
\end{figure}
\begin{figure}[htbp]
\centering
\includegraphics[width=14.5cm]{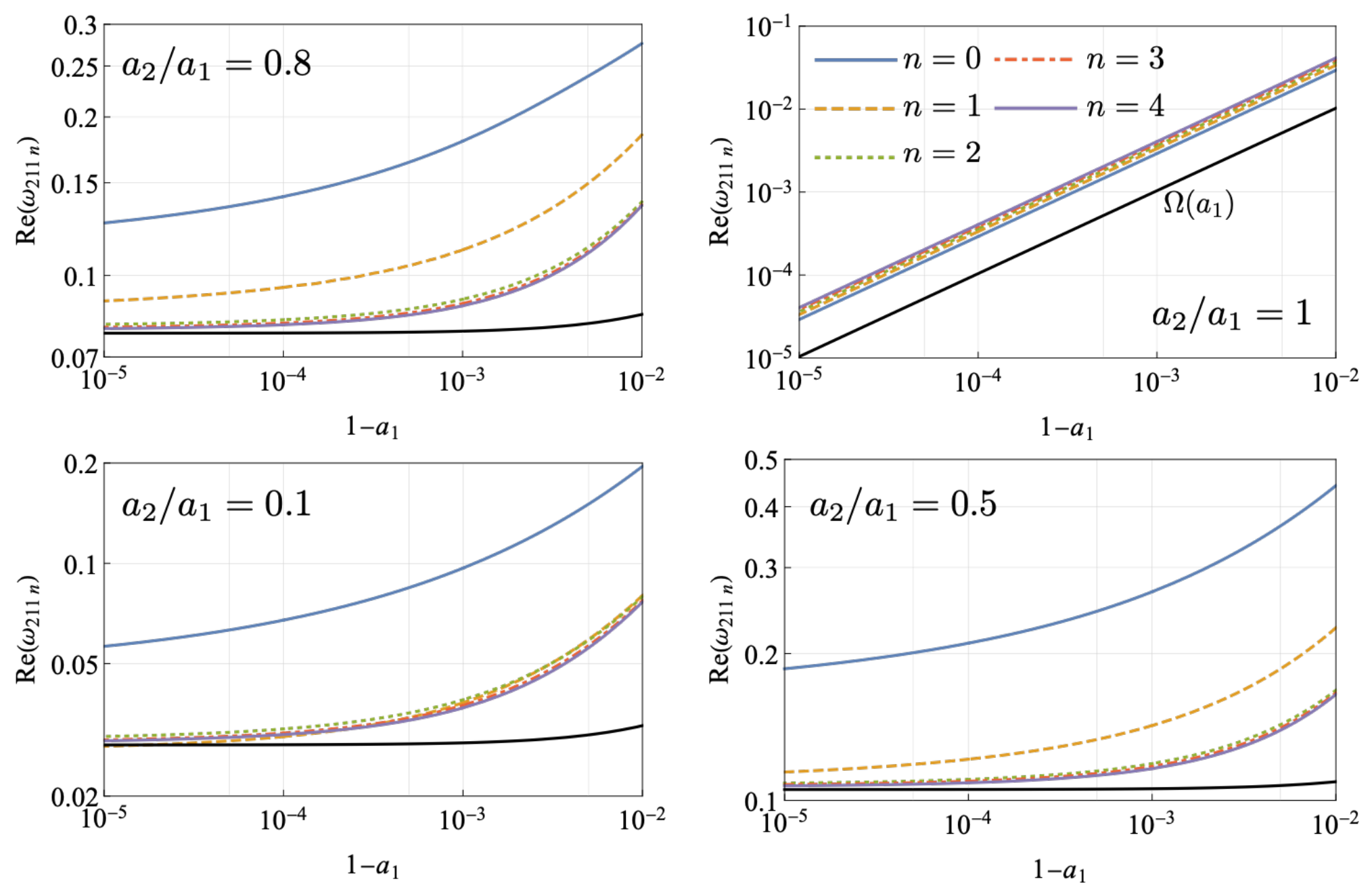}
\caption{Real part of the QN frequencies in the ultra-spinning limit. The spin ratio is fixed as $a_2/a_1 = 0.1$, $0.5$, $0.8$, and $1$, the mass parameter is $M=10$, and the harmonic mode is $(l,m_1,m_2) = (2,1,1)$. The QN frequencies are shown up to the 4th overtone. For the equal spins, the real part of QN frequencies approaches zero as $\Omega \to 0$ in the ultra-spinning limit. For unequal spin case, $\text{Re} (\omega_{211n})$ saturates at $\Omega > 0$. The black lines indicate the value of $\Omega$.}
\label{unequal_real_qnf}
\end{figure}

We have numerically shown that the damping rates of the type-II QN modes of a small \kdsf black hole with unequal spins also have the periodicity of $2 \pi T_{\rm H}$. We here numerically check that the intriguing thermodynamic property of highly damped modes holds even for \kdsf black holes with large mass and unequal spins.
We show the separation $\Delta \text{Im} (\omega_{lm_1m_2n})$ as a function of $1-a_1$ in Figure \ref{unequal_Hawking_temp}. For highly damped modes, the separation approaches $2 \pi T_{\rm H}$ in the ultra-spinning limit. Also, Figure \ref{unequal_real_qnf} shows the real part of the type-II QN frequencies for up to the 4th overtone, and one can see that $\text{Re} (\omega_{lm_1m_2n})$ approaches $\text{Re} (\omega) = \Omega$ as $a_1 \to 1$. In the next section, we discuss an implication we can obtain by combining our result and the Hod's conjecture on the black hole area quantization \cite{Hod:1998vk}.

\section{Area quantization of the \kdsf black hole}
\label{sec:areaQ}
\begin{figure}[t]
\centering
\includegraphics[width=10cm]{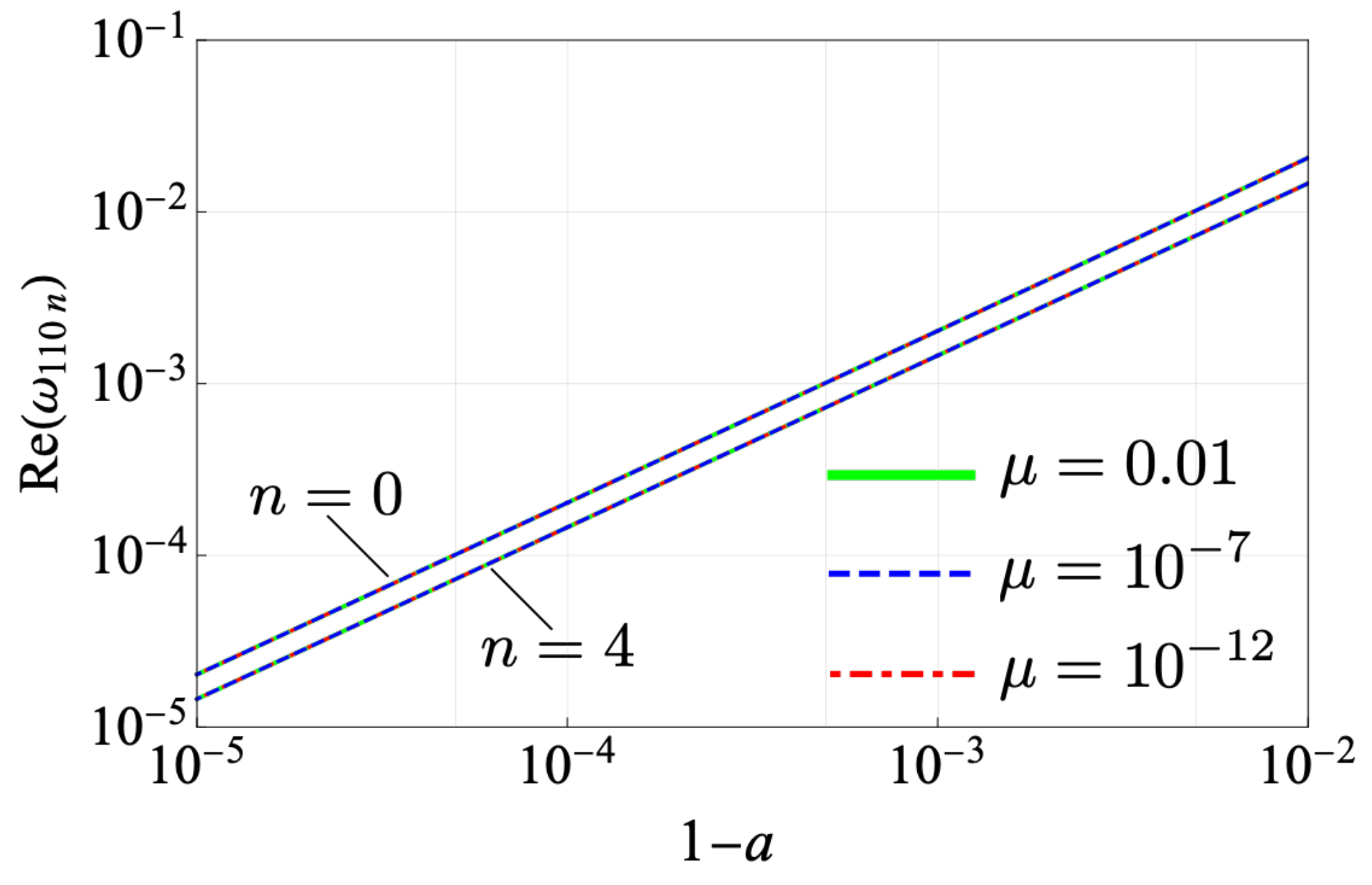}
\caption{Real part of QN frequencies of $(l,m_1,m_2) = (1,1,0)$ and $n=0$ and $4$. The mass of the scalar field is set to $\mu = 0.01$, $10^{-7}$, and $10^{-12}$. The parameters are set to $M=10$ and $a_1=a_2$.}
\label{mu_depend}
\end{figure}

In the previous section, we found that the real part of the highly damped QN frequencies approaches the superradiant frequency.
The strong localization at $\text{Re} (\omega) = \Omega$ can be seen in the low-temperature or ultra-spinning limits. According to the Hod's conjecture \cite{Hod:1998vk}, the asymptotic value of the real part of QN frequencies of massless fields determines the smallest size of a quantized horizon area, $\Delta A$, and the horizon area is given by $A = N_{\rm max} \Delta A$. The integer $N_{\rm max}$ is the total number of the unit area on the horizon. In our numerical code, it is technically difficult to handle the exactly massless scalar perturbation. On the other hand, at least in the range of $10^{-12} \leq \mu \leq 0.01$, we confirm that the values of QN frequencies converge well (see Figure \ref{mu_depend}). 
Hence we assume that the scalar field with $\mu = 0.01$ is effectively massless in our situation. In this section, we discuss the relation between the QN modes and the horizon area quantization for the \kdsf black holes.

Let us briefly review the Hod's conjecture and its physical interpretation. Hod proposed that the overtone number $n$ may be interpreted as a quantum number characterizing the energy levels of a black hole as an analogy of a hydrogen atom, as in the Bohr's corresponding principle. Based on this idea, the discretized values of the real part of QN frequencies might be associated with the energy of quanta which the black hole can emit or absorb. From this point of view, the mass of a Schwarzschild black hole may be quantized as
\begin{equation}
\Delta M = \lim_{n \to \infty} \text{Re} (\omega_{lm_1m_2n}),
\label{delm_eq_wlmmn}
\end{equation}
which leads to the horizon area quantization as $\Delta A/4G = \Delta M/T_{\rm H}$.
At least, for the Schwarzschild \cite{Motl:2002hd} and Kerr \cite{Berti:2003jh,Berti:2004um} black holes, the asymptotic value of the real part of QN frequencies is independent of the angular index $l$ and is universal. The quantization of the \kdsf black hole mass may also lead to the horizon area quantization\footnote{However, the Hod's conjecture may be subtle for the Kerr spacetime if the asymptotic value of the real part of QN frequencies is $m \Omega$ for a Kerr black hole with ($a> 0$). The naive application of the Hod's conjecture to that case leads to $\Delta A = 0$. See Ref. \cite{Berti:2003jh,Berti:2004um} for more details.}. The first law of black hole thermodynamics in the \kdsf spacetime is \cite{Gibbons:2004ai}
\begin{equation}\displaystyle
\Delta {\cal M} = T_{\rm H} \frac{\Delta A}{4G} + \sum_{i=1}^2 \Omega_i \Delta J_i.
\end{equation}
Assuming (\ref{delm_eq_wlmmn}), $\Delta J_i = m_i$, and 
\begin{equation}
\lim_{n \to \infty} \text{Re} (\omega_{lm_1m_2n}) \to \Omega = \sum_{i} m_i \Omega_i,
\label{assumption_SRF}
\end{equation}
the first law reduces to
\begin{equation}
T_{\rm H} \frac{\Delta A}{4G} = 0.
\label{zero_area}
\end{equation}
It means that when $T_{\rm H} \neq 0$, the horizon is no longer quantized and becomes a {\textit continuum}, at least based on the Hod's conjecture. Note that our numerical results just imply that the real part of QN frequencies for the first several overtones approaches $\Omega$, in the low-temperature limit. Therefore, the strong convergence of $\text{Re} (\omega_{lm_1m_2n})$ to $\Omega$ for $n \to \infty$ is still non-trivial from our investigation. Also, for small black holes ${\cal M} \ll 1$, the lowest Hawking temperature is zero, and so it is non-trivial if $\Delta A = 0$ at $T_{\rm H} = 0$. On the other hand, for large black holes, the temperature is non-zero and finite even at the ultra-spinning limit, and therefore, the assumption of (\ref{assumption_SRF}) leads to $\Delta A = 0$ for ${\cal M} \gtrsim 1$.

We can show that in the ultra-spinning limit, the horizon area of the \kdsf black hole becomes singular. To demonstrate this, let us perform the following coordinate transformations in the metric (\ref{metric}):
\begin{equation}
\phi \to \sqrt{1-a_1^2} \phi, \ \psi \to \sqrt{1-a_2^2} \psi.
\end{equation}
Considering the geometry of constant $(t,r)$ surface in the near-horizon limit ($r \to r_+$), the line element for the ultra-spinning limit reduces to
\begin{equation}
ds^2 = \lim_{a \to 1} \frac{r_+^2+1}{1-a^2} \left[d\theta^2 + \frac{(r_+^2+1)}{r_+^2} \left( \sin^2 \theta d\phi + \cos^2 \theta d \psi\right)^2 \right],
\end{equation}
where we take $a_1=a_2=a$. This shows that in the limit, the horizon area diverges in the coordinates.
On the other hand, when we take the limit of $a_1 \to 1$ while keeping $a_2 < 1$, the topology of the horizon becomes non-compact. As a simple example, let us look into the case of $a_1 \to 1$ and $a_2= 0$. In that case, the metric reduces to
\begin{align}
ds^2= \frac{(r_+^2+1)^2}{\rho^2} \sin^4 \theta d\phi^2 + \frac{r_+^4}{\rho^2} \sin^2 \theta \cos^2 \theta d\psi^2+ \frac{r_+^2 (r_+^2+1)}{\rho^2} \cos^4 \theta d \psi^2 
+ \frac{\rho^2}{\sin^2 \theta} d\theta^2,
\label{horizon_metric_2}
\end{align}
where we performed the following transformations for the original metric
\begin{equation}
\phi \to (1-a_1^2) \phi, \ \psi \to (1-a_2^2) \psi.
\end{equation}
The metric in (\ref{horizon_metric_2}) appears to be ill-defined at $\theta = 0$. However, by performing a further transformation of
\begin{equation}
\xi = 1- \cos \theta,
\end{equation}
the metric near the pole of $\theta = 0$ is
\begin{equation}
ds^2 \simeq (r_+^2+1) \left(4 \xi^2 d \phi^2 + \frac{d\xi^2}{4 \xi^2} \right) + r_+^2 d\psi^2,
\end{equation}
which has the submanifold of $\mathbb{H}^2$ on $\phi-\xi$ surface, and we find that the constant $(t,r)$ surface is non-compact in the ultra-spinning limit. One can easily find that for $a_1 \to 1$ and $a_2 \neq 0$, the $\phi-\xi$ surface near $r= r_+$ and $\theta = 0$ is also non-compact as its metric is 
\begin{equation}
ds^2 \simeq (r_+^2+1) \left( 4 \xi^2 \frac{r_+^2+a_2^2}{r_+^2} d\phi^2 + \frac{1}{1-a_2^2} \frac{d\xi^2}{4 \xi^2} \right).
\end{equation}
Indeed, a similar non-compact horizon appears even in the Kerr-AdS${}_4$ black hole \cite{Klemm:2014rda,Hennigar:2014cfa,Hennigar:2015cja}. 

In summary, we have applied the Hod's proposal to our result that the real part of scalar QN frequencies approaches the superradiant frequency in the low-temperature or ultra-spinning limits.
Then we have concluded that it leads to a continuous horizon area, i.e. unquantized horizon area, when naively applying the Hod's conjecture to \kdsf black holes. Also, in the special cases involving the ultra-spinning regime, the total horizon area diverges, or the topology of the \kdsf black hole horizon is non-compact.

\section{Discussion and Conclusion}
\label{sec:conclusion}
In this paper, we have numerically investigated the structure of scalar quasi-normal (QN) frequencies of the five-dimensional Kerr anti-de Sitter (Kerr-AdS${}_5$) black hole. Our numerical investigation covers a broad range of parameter regions for the \kdsf black hole.

In Sec. \ref{sec:equal_small_BH}, we have studied the instability of small \kdsf black holes with equal spins. We have confirmed that the strong instability is caused by $l=1$, and the most unstable QN mode has its frequency close to the superradiant frequency defined by $\Omega \equiv m_1 \Omega_{+,1} + m_2 \Omega_{+,2}$. This is the case even for unequal spins, as shown in Sec. \ref{sec:unequal_small}. 
We have also investigated if the superradiant instability is amplified when the \kdsf black hole has a hierarchy between the two spins, i.e., $a_2/a_1 < 1$. As we have checked that the most unstable harmonics is $(l,m_1,m_2) = (1,1,0)$ for $a_2 < a_1$ (Figure \ref{ell_neq}), we computed the unstable QN frequencies of $(1,1,0)$ mode for different spin ratios while fixing the angular velocity $\Omega_1$ which is equivalent to the superradiant frequency for $(1,1,0)$ mode. 
Then, we have found that the stronger the hierarchy of the two spins is, the more enhanced the instability is (Figure \ref{QNM_neq}). We conclude that the superradiant instability of the \kdsf black holes is enhanced when the symmetry of the spacetime is reduced. The QN modes that could induce the instability of small \kdsf black holes localize near the real axis in the complex frequency plane.
We call those modes type-I modes. On the other hand, we have investigated highly damped QN modes (type-II modes), which differ from the type-I modes and might be relevant to the thermodynamic nature of the \kdsf black hole as the separation of their imaginary part is $\sim 2 \pi T_{\rm H}$. We confirmed that the separation matches $2 \pi T_{\rm H}$ in high accuracy in the low-temperature limit.

In Sec. \ref{sec:equal_large_BH} and Sec. \ref{sec:unequal_large}, we have investigated the scalar QN modes for large \kdsf black holes with equal and unequal spins, respectively. As spin parameters approach the AdS curvature radius, that is the ultra-spinning limit ($a_i \to 1$), type-I modes are suppressed and type-II modes are excited along $\text{Re} (\omega) = \Omega$. 
The type-II modes with higher damping rates have the periodic separation in their imaginary parts, and the separation for the first several tones matches the surface gravity of the horizon when $a_i \to 1$ and $\mu \ll 1$ (see Figure \ref{FIGa}). The thermal nature holds even for large \kdsf black holes with unequal spins (Figure \ref{unequal_Hawking_temp} and \ref{unequal_real_qnf}). It would be interesting to study how this property can be relevant to the pole structure of the thermal Green's function in the corresponding conformal field theory (CFT) on the AdS boundary.

Based on the Hod's conjecture regarding the horizon area quantization \cite{Hod:1998vk}, the asymptotic value of $\text{Re} (\omega_{lm_1m_2n})$ plays an important role in determining the {\it one-bit size} of a black hole area $\Delta A$ as $T_{\rm H} \Delta A/4G = \lim_{n \to \infty} \text{Re} (\omega_{lm_1m_2n}) - \Omega$. Therefore, as we discussed in Sec. \ref{sec:areaQ}, the convergence of the real part of type-II QN frequencies to $\Omega$ for large $n$ implies that the horizon area is no longer quantized but is continuous when naively applying the Hod's conjecture to the \kdsf black holes. From our numerical result, the convergence is likely at least for the low-temperature and ultra-spinning regimes. We leave the further study of highly damped QN modes in an analytical way for future work. We also analyzed the horizon topology of the \kdsf black hole and found that in general, an ultra-spinning \kdsf black hole has its non-compact horizon.

\section*{Acknowledgement}
This work was supported in part by the Special Postdoctoral Researcher (SPDR) Program at RIKEN (NO), Incentive Research Project at RIKEN (NO), Grant-in-Aid for Scientific Research (KAKENHI) project (21K20371) (NO), and Japan Society for the Promotion of Science(JSPS) Grant No. 20J22946(KU).


\begin{thebibliography}{37}%
\makeatletter
\providecommand \@ifxundefined [1]{%
 \@ifx{#1\undefined}
}%
\providecommand \@ifnum [1]{%
 \ifnum #1\expandafter \@firstoftwo
 \else \expandafter \@secondoftwo
 \fi
}%
\providecommand \@ifx [1]{%
 \ifx #1\expandafter \@firstoftwo
 \else \expandafter \@secondoftwo
 \fi
}%
\providecommand \natexlab [1]{#1}%
\providecommand \enquote  [1]{``#1''}%
\providecommand \bibnamefont  [1]{#1}%
\providecommand \bibfnamefont [1]{#1}%
\providecommand \citenamefont [1]{#1}%
\providecommand \href@noop [0]{\@secondoftwo}%
\providecommand \href [0]{\begingroup \@sanitize@url \@href}%
\providecommand \@href[1]{\@@startlink{#1}\@@href}%
\providecommand \@@href[1]{\endgroup#1\@@endlink}%
\providecommand \@sanitize@url [0]{\catcode `\\12\catcode `\$12\catcode
  `\&12\catcode `\#12\catcode `\^12\catcode `\_12\catcode `\%12\relax}%
\providecommand \@@startlink[1]{}%
\providecommand \@@endlink[0]{}%
\providecommand \url  [0]{\begingroup\@sanitize@url \@url }%
\providecommand \@url [1]{\endgroup\@href {#1}{\urlprefix }}%
\providecommand \urlprefix  [0]{URL }%
\providecommand \Eprint [0]{\href }%
\providecommand \doibase [0]{http://dx.doi.org/}%
\providecommand \selectlanguage [0]{\@gobble}%
\providecommand \bibinfo  [0]{\@secondoftwo}%
\providecommand \bibfield  [0]{\@secondoftwo}%
\providecommand \translation [1]{[#1]}%
\providecommand \BibitemOpen [0]{}%
\providecommand \bibitemStop [0]{}%
\providecommand \bibitemNoStop [0]{.\EOS\space}%
\providecommand \EOS [0]{\spacefactor3000\relax}%
\providecommand \BibitemShut  [1]{\csname bibitem#1\endcsname}%
\let\auto@bib@innerbib\@empty
\bibitem [{\citenamefont {'t~Hooft}(1993)}]{tHooft:1993dmi}%
  \BibitemOpen
  \bibfield  {author} {\bibinfo {author} {\bibfnamefont {G.}~\bibnamefont
  {'t~Hooft}},\ }\href@noop {} {\bibfield  {journal} {\bibinfo  {journal}
  {Conf. Proc. C}\ }\textbf {\bibinfo {volume} {930308}},\ \bibinfo {pages}
  {284} (\bibinfo {year} {1993})},\ \Eprint
  {http://arxiv.org/abs/gr-qc/9310026} {arXiv:gr-qc/9310026} \BibitemShut
  {NoStop}%
\bibitem [{\citenamefont {Susskind}, \citenamefont {Thorlacius},\ and\
  \citenamefont {Uglum}(1993)}]{Susskind:1993if}%
  \BibitemOpen
  \bibfield  {author} {\bibinfo {author} {\bibfnamefont {L.}~\bibnamefont
  {Susskind}}, \bibinfo {author} {\bibfnamefont {L.}~\bibnamefont
  {Thorlacius}}, \ and\ \bibinfo {author} {\bibfnamefont {J.}~\bibnamefont
  {Uglum}},\ }\href {\doibase 10.1103/PhysRevD.48.3743} {\bibfield  {journal}
  {\bibinfo  {journal} {Phys. Rev. D}\ }\textbf {\bibinfo {volume} {48}},\
  \bibinfo {pages} {3743} (\bibinfo {year} {1993})},\ \Eprint
  {http://arxiv.org/abs/hep-th/9306069} {arXiv:hep-th/9306069} \BibitemShut
  {NoStop}%
\bibitem [{\citenamefont {Susskind}(1995)}]{Susskind:1994vu}%
  \BibitemOpen
  \bibfield  {author} {\bibinfo {author} {\bibfnamefont {L.}~\bibnamefont
  {Susskind}},\ }\href {\doibase 10.1063/1.531249} {\bibfield  {journal}
  {\bibinfo  {journal} {J. Math. Phys.}\ }\textbf {\bibinfo {volume} {36}},\
  \bibinfo {pages} {6377} (\bibinfo {year} {1995})},\ \Eprint
  {http://arxiv.org/abs/hep-th/9409089} {arXiv:hep-th/9409089} \BibitemShut
  {NoStop}%
\bibitem [{\citenamefont {Maldacena}(1998)}]{Maldacena:1997re}%
  \BibitemOpen
  \bibfield  {author} {\bibinfo {author} {\bibfnamefont {J.~M.}\ \bibnamefont
  {Maldacena}},\ }\href {\doibase 10.1023/A:1026654312961} {\bibfield
  {journal} {\bibinfo  {journal} {Adv. Theor. Math. Phys.}\ }\textbf {\bibinfo
  {volume} {2}},\ \bibinfo {pages} {231} (\bibinfo {year} {1998})},\ \Eprint
  {http://arxiv.org/abs/hep-th/9711200} {arXiv:hep-th/9711200} \BibitemShut
  {NoStop}%
\bibitem [{\citenamefont {Horava}\ and\ \citenamefont
  {Witten}(1996)}]{Horava:1996ma}%
  \BibitemOpen
  \bibfield  {author} {\bibinfo {author} {\bibfnamefont {P.}~\bibnamefont
  {Horava}}\ and\ \bibinfo {author} {\bibfnamefont {E.}~\bibnamefont
  {Witten}},\ }\href {\doibase 10.1016/0550-3213(96)00308-2} {\bibfield
  {journal} {\bibinfo  {journal} {Nucl. Phys. B}\ }\textbf {\bibinfo {volume}
  {475}},\ \bibinfo {pages} {94} (\bibinfo {year} {1996})},\ \Eprint
  {http://arxiv.org/abs/hep-th/9603142} {arXiv:hep-th/9603142} \BibitemShut
  {NoStop}%
\bibitem [{\citenamefont {Arkani-Hamed}, \citenamefont {Dimopoulos},\ and\
  \citenamefont {Dvali}(1998)}]{Arkani-Hamed:1998jmv}%
  \BibitemOpen
  \bibfield  {author} {\bibinfo {author} {\bibfnamefont {N.}~\bibnamefont
  {Arkani-Hamed}}, \bibinfo {author} {\bibfnamefont {S.}~\bibnamefont
  {Dimopoulos}}, \ and\ \bibinfo {author} {\bibfnamefont {G.~R.}\ \bibnamefont
  {Dvali}},\ }\href {\doibase 10.1016/S0370-2693(98)00466-3} {\bibfield
  {journal} {\bibinfo  {journal} {Phys. Lett. B}\ }\textbf {\bibinfo {volume}
  {429}},\ \bibinfo {pages} {263} (\bibinfo {year} {1998})},\ \Eprint
  {http://arxiv.org/abs/hep-ph/9803315} {arXiv:hep-ph/9803315} \BibitemShut
  {NoStop}%
\bibitem [{\citenamefont {Randall}\ and\ \citenamefont
  {Sundrum}(1999{\natexlab{a}})}]{Randall:1999vf}%
  \BibitemOpen
  \bibfield  {author} {\bibinfo {author} {\bibfnamefont {L.}~\bibnamefont
  {Randall}}\ and\ \bibinfo {author} {\bibfnamefont {R.}~\bibnamefont
  {Sundrum}},\ }\href {\doibase 10.1103/PhysRevLett.83.4690} {\bibfield
  {journal} {\bibinfo  {journal} {Phys. Rev. Lett.}\ }\textbf {\bibinfo
  {volume} {83}},\ \bibinfo {pages} {4690} (\bibinfo {year}
  {1999}{\natexlab{a}})},\ \Eprint {http://arxiv.org/abs/hep-th/9906064}
  {arXiv:hep-th/9906064} \BibitemShut {NoStop}%
\bibitem [{\citenamefont {Randall}\ and\ \citenamefont
  {Sundrum}(1999{\natexlab{b}})}]{Randall:1999ee}%
  \BibitemOpen
  \bibfield  {author} {\bibinfo {author} {\bibfnamefont {L.}~\bibnamefont
  {Randall}}\ and\ \bibinfo {author} {\bibfnamefont {R.}~\bibnamefont
  {Sundrum}},\ }\href {\doibase 10.1103/PhysRevLett.83.3370} {\bibfield
  {journal} {\bibinfo  {journal} {Phys. Rev. Lett.}\ }\textbf {\bibinfo
  {volume} {83}},\ \bibinfo {pages} {3370} (\bibinfo {year}
  {1999}{\natexlab{b}})},\ \Eprint {http://arxiv.org/abs/hep-ph/9905221}
  {arXiv:hep-ph/9905221} \BibitemShut {NoStop}%
\bibitem [{\citenamefont {Arkani-Hamed}\ \emph {et~al.}(2000)\citenamefont
  {Arkani-Hamed}, \citenamefont {Dimopoulos}, \citenamefont {Dvali},\ and\
  \citenamefont {Kaloper}}]{Arkani-Hamed:1999wga}%
  \BibitemOpen
  \bibfield  {author} {\bibinfo {author} {\bibfnamefont {N.}~\bibnamefont
  {Arkani-Hamed}}, \bibinfo {author} {\bibfnamefont {S.}~\bibnamefont
  {Dimopoulos}}, \bibinfo {author} {\bibfnamefont {G.~R.}\ \bibnamefont
  {Dvali}}, \ and\ \bibinfo {author} {\bibfnamefont {N.}~\bibnamefont
  {Kaloper}},\ }\href {\doibase 10.1103/PhysRevLett.84.586} {\bibfield
  {journal} {\bibinfo  {journal} {Phys. Rev. Lett.}\ }\textbf {\bibinfo
  {volume} {84}},\ \bibinfo {pages} {586} (\bibinfo {year} {2000})},\ \Eprint
  {http://arxiv.org/abs/hep-th/9907209} {arXiv:hep-th/9907209} \BibitemShut
  {NoStop}%
\bibitem [{\citenamefont {Susskind}(2003)}]{Susskind:2003kw}%
  \BibitemOpen
  \bibfield  {author} {\bibinfo {author} {\bibfnamefont {L.}~\bibnamefont
  {Susskind}},\ }\href@noop {} {\ ,\ \bibinfo {pages} {247} (\bibinfo {year}
  {2003})},\ \Eprint {http://arxiv.org/abs/hep-th/0302219}
  {arXiv:hep-th/0302219} \BibitemShut {NoStop}%
\bibitem [{\citenamefont {Witten}(1998)}]{Witten:1998qj}%
  \BibitemOpen
  \bibfield  {author} {\bibinfo {author} {\bibfnamefont {E.}~\bibnamefont
  {Witten}},\ }\href {\doibase 10.4310/ATMP.1998.v2.n2.a2} {\bibfield
  {journal} {\bibinfo  {journal} {Adv. Theor. Math. Phys.}\ }\textbf {\bibinfo
  {volume} {2}},\ \bibinfo {pages} {253} (\bibinfo {year} {1998})},\ \Eprint
  {http://arxiv.org/abs/hep-th/9802150} {arXiv:hep-th/9802150} \BibitemShut
  {NoStop}%
\bibitem [{\citenamefont {Banks}\ \emph {et~al.}(1998)\citenamefont {Banks},
  \citenamefont {Douglas}, \citenamefont {Horowitz},\ and\ \citenamefont
  {Martinec}}]{Banks:1998dd}%
  \BibitemOpen
  \bibfield  {author} {\bibinfo {author} {\bibfnamefont {T.}~\bibnamefont
  {Banks}}, \bibinfo {author} {\bibfnamefont {M.~R.}\ \bibnamefont {Douglas}},
  \bibinfo {author} {\bibfnamefont {G.~T.}\ \bibnamefont {Horowitz}}, \ and\
  \bibinfo {author} {\bibfnamefont {E.~J.}\ \bibnamefont {Martinec}},\
  }\href@noop {} {\  (\bibinfo {year} {1998})},\ \Eprint
  {http://arxiv.org/abs/hep-th/9808016} {arXiv:hep-th/9808016} \BibitemShut
  {NoStop}%
\bibitem [{\citenamefont {Kraus}(1999)}]{Kraus:1999it}%
  \BibitemOpen
  \bibfield  {author} {\bibinfo {author} {\bibfnamefont {P.}~\bibnamefont
  {Kraus}},\ }\href {\doibase 10.1088/1126-6708/1999/12/011} {\bibfield
  {journal} {\bibinfo  {journal} {JHEP}\ }\textbf {\bibinfo {volume} {12}},\
  \bibinfo {pages} {011} (\bibinfo {year} {1999})},\ \Eprint
  {http://arxiv.org/abs/hep-th/9910149} {arXiv:hep-th/9910149} \BibitemShut
  {NoStop}%
\bibitem [{\citenamefont {Ida}(2000)}]{Ida:1999ui}%
  \BibitemOpen
  \bibfield  {author} {\bibinfo {author} {\bibfnamefont {D.}~\bibnamefont
  {Ida}},\ }\href {\doibase 10.1088/1126-6708/2000/09/014} {\bibfield
  {journal} {\bibinfo  {journal} {JHEP}\ }\textbf {\bibinfo {volume} {09}},\
  \bibinfo {pages} {014} (\bibinfo {year} {2000})},\ \Eprint
  {http://arxiv.org/abs/gr-qc/9912002} {arXiv:gr-qc/9912002} \BibitemShut
  {NoStop}%
\bibitem [{\citenamefont {Mukohyama}(2000)}]{Mukohyama:1999qx}%
  \BibitemOpen
  \bibfield  {author} {\bibinfo {author} {\bibfnamefont {S.}~\bibnamefont
  {Mukohyama}},\ }\href {\doibase 10.1016/S0370-2693(99)01505-1} {\bibfield
  {journal} {\bibinfo  {journal} {Phys. Lett. B}\ }\textbf {\bibinfo {volume}
  {473}},\ \bibinfo {pages} {241} (\bibinfo {year} {2000})},\ \Eprint
  {http://arxiv.org/abs/hep-th/9911165} {arXiv:hep-th/9911165} \BibitemShut
  {NoStop}%
\bibitem [{\citenamefont {Birmingham}, \citenamefont {Sachs},\ and\
  \citenamefont {Solodukhin}(2002)}]{Birmingham:2001pj}%
  \BibitemOpen
  \bibfield  {author} {\bibinfo {author} {\bibfnamefont {D.}~\bibnamefont
  {Birmingham}}, \bibinfo {author} {\bibfnamefont {I.}~\bibnamefont {Sachs}}, \
  and\ \bibinfo {author} {\bibfnamefont {S.~N.}\ \bibnamefont {Solodukhin}},\
  }\href {\doibase 10.1103/PhysRevLett.88.151301} {\bibfield  {journal}
  {\bibinfo  {journal} {Phys. Rev. Lett.}\ }\textbf {\bibinfo {volume} {88}},\
  \bibinfo {pages} {151301} (\bibinfo {year} {2002})},\ \Eprint
  {http://arxiv.org/abs/hep-th/0112055} {arXiv:hep-th/0112055} \BibitemShut
  {NoStop}%
\bibitem [{\citenamefont {Hod}(1998)}]{Hod:1998vk}%
  \BibitemOpen
  \bibfield  {author} {\bibinfo {author} {\bibfnamefont {S.}~\bibnamefont
  {Hod}},\ }\href {\doibase 10.1103/PhysRevLett.81.4293} {\bibfield  {journal}
  {\bibinfo  {journal} {Phys. Rev. Lett.}\ }\textbf {\bibinfo {volume} {81}},\
  \bibinfo {pages} {4293} (\bibinfo {year} {1998})},\ \Eprint
  {http://arxiv.org/abs/gr-qc/9812002} {arXiv:gr-qc/9812002} \BibitemShut
  {NoStop}%
\bibitem [{\citenamefont {Myers}\ and\ \citenamefont
  {Perry}(1986)}]{Myers:1986un}%
  \BibitemOpen
  \bibfield  {author} {\bibinfo {author} {\bibfnamefont {R.~C.}\ \bibnamefont
  {Myers}}\ and\ \bibinfo {author} {\bibfnamefont {M.~J.}\ \bibnamefont
  {Perry}},\ }\href {\doibase 10.1016/0003-4916(86)90186-7} {\bibfield
  {journal} {\bibinfo  {journal} {Annals Phys.}\ }\textbf {\bibinfo {volume}
  {172}},\ \bibinfo {pages} {304} (\bibinfo {year} {1986})}\BibitemShut
  {NoStop}%
\bibitem [{\citenamefont {Aliev}\ and\ \citenamefont
  {Delice}(2009)}]{Aliev:2008yk}%
  \BibitemOpen
  \bibfield  {author} {\bibinfo {author} {\bibfnamefont {A.~N.}\ \bibnamefont
  {Aliev}}\ and\ \bibinfo {author} {\bibfnamefont {O.}~\bibnamefont {Delice}},\
  }\href {\doibase 10.1103/PhysRevD.79.024013} {\bibfield  {journal} {\bibinfo
  {journal} {Phys. Rev. D}\ }\textbf {\bibinfo {volume} {79}},\ \bibinfo
  {pages} {024013} (\bibinfo {year} {2009})},\ \Eprint
  {http://arxiv.org/abs/0808.0280} {arXiv:0808.0280 [hep-th]} \BibitemShut
  {NoStop}%
\bibitem [{\citenamefont {Murata}(2009)}]{Murata:2008xr}%
  \BibitemOpen
  \bibfield  {author} {\bibinfo {author} {\bibfnamefont {K.}~\bibnamefont
  {Murata}},\ }\href {\doibase 10.1143/PTP.121.1099} {\bibfield  {journal}
  {\bibinfo  {journal} {Prog. Theor. Phys.}\ }\textbf {\bibinfo {volume}
  {121}},\ \bibinfo {pages} {1099} (\bibinfo {year} {2009})},\ \Eprint
  {http://arxiv.org/abs/0812.0718} {arXiv:0812.0718 [hep-th]} \BibitemShut
  {NoStop}%
\bibitem [{\citenamefont {Murata}, \citenamefont {Nishioka},\ and\
  \citenamefont {Tanahashi}(2009)}]{Murata:2009jt}%
  \BibitemOpen
  \bibfield  {author} {\bibinfo {author} {\bibfnamefont {K.}~\bibnamefont
  {Murata}}, \bibinfo {author} {\bibfnamefont {T.}~\bibnamefont {Nishioka}}, \
  and\ \bibinfo {author} {\bibfnamefont {N.}~\bibnamefont {Tanahashi}},\ }\href
  {\doibase 10.1143/PTP.121.941} {\bibfield  {journal} {\bibinfo  {journal}
  {Prog. Theor. Phys.}\ }\textbf {\bibinfo {volume} {121}},\ \bibinfo {pages}
  {941} (\bibinfo {year} {2009})},\ \Eprint {http://arxiv.org/abs/0901.2574}
  {arXiv:0901.2574 [hep-th]} \BibitemShut {NoStop}%
\bibitem [{\citenamefont {Cardoso}\ \emph {et~al.}(2014)\citenamefont
  {Cardoso}, \citenamefont {Dias}, \citenamefont {Hartnett}, \citenamefont
  {Lehner},\ and\ \citenamefont {Santos}}]{Cardoso:2013pza}%
  \BibitemOpen
  \bibfield  {author} {\bibinfo {author} {\bibfnamefont {V.}~\bibnamefont
  {Cardoso}}, \bibinfo {author} {\bibfnamefont {O.~J.~C.}\ \bibnamefont
  {Dias}}, \bibinfo {author} {\bibfnamefont {G.~S.}\ \bibnamefont {Hartnett}},
  \bibinfo {author} {\bibfnamefont {L.}~\bibnamefont {Lehner}}, \ and\ \bibinfo
  {author} {\bibfnamefont {J.~E.}\ \bibnamefont {Santos}},\ }\href {\doibase
  10.1007/JHEP04(2014)183} {\bibfield  {journal} {\bibinfo  {journal} {JHEP}\
  }\textbf {\bibinfo {volume} {04}},\ \bibinfo {pages} {183} (\bibinfo {year}
  {2014})},\ \Eprint {http://arxiv.org/abs/1312.5323} {arXiv:1312.5323
  [hep-th]} \BibitemShut {NoStop}%
\bibitem [{\citenamefont {Aref'eva}, \citenamefont {Golubtsova},\ and\
  \citenamefont {Gourgoulhon}(2021)}]{Arefeva:2020jvo}%
  \BibitemOpen
  \bibfield  {author} {\bibinfo {author} {\bibfnamefont {I.~Y.}\ \bibnamefont
  {Aref'eva}}, \bibinfo {author} {\bibfnamefont {A.~A.}\ \bibnamefont
  {Golubtsova}}, \ and\ \bibinfo {author} {\bibfnamefont {E.}~\bibnamefont
  {Gourgoulhon}},\ }\href {\doibase 10.1007/JHEP04(2021)169} {\bibfield
  {journal} {\bibinfo  {journal} {JHEP}\ }\textbf {\bibinfo {volume} {04}},\
  \bibinfo {pages} {169} (\bibinfo {year} {2021})},\ \Eprint
  {http://arxiv.org/abs/2004.12984} {arXiv:2004.12984 [hep-th]} \BibitemShut
  {NoStop}%
\bibitem [{\citenamefont {Nian}\ and\ \citenamefont
  {Pando~Zayas}(2020)}]{Nian:2020qsk}%
  \BibitemOpen
  \bibfield  {author} {\bibinfo {author} {\bibfnamefont {J.}~\bibnamefont
  {Nian}}\ and\ \bibinfo {author} {\bibfnamefont {L.~A.}\ \bibnamefont
  {Pando~Zayas}},\ }\href {\doibase 10.1007/JHEP07(2020)120} {\bibfield
  {journal} {\bibinfo  {journal} {JHEP}\ }\textbf {\bibinfo {volume} {07}},\
  \bibinfo {pages} {120} (\bibinfo {year} {2020})},\ \Eprint
  {http://arxiv.org/abs/2003.02770} {arXiv:2003.02770 [hep-th]} \BibitemShut
  {NoStop}%
\bibitem [{\citenamefont {Barrag\'an~Amado}, \citenamefont {Carneiro
  Da~Cunha},\ and\ \citenamefont {Pallante}(2019)}]{BarraganAmado:2018zpa}%
  \BibitemOpen
  \bibfield  {author} {\bibinfo {author} {\bibfnamefont {J.}~\bibnamefont
  {Barrag\'an~Amado}}, \bibinfo {author} {\bibfnamefont {B.}~\bibnamefont
  {Carneiro Da~Cunha}}, \ and\ \bibinfo {author} {\bibfnamefont
  {E.}~\bibnamefont {Pallante}},\ }\href {\doibase 10.1103/PhysRevD.99.105006}
  {\bibfield  {journal} {\bibinfo  {journal} {Phys. Rev. D}\ }\textbf {\bibinfo
  {volume} {99}},\ \bibinfo {pages} {105006} (\bibinfo {year} {2019})},\
  \Eprint {http://arxiv.org/abs/1812.08921} {arXiv:1812.08921 [hep-th]}
  \BibitemShut {NoStop}%
\bibitem [{\citenamefont {Amado}, \citenamefont {da~Cunha},\ and\ \citenamefont
  {Pallante}(2021)}]{Amado:2021erf}%
  \BibitemOpen
  \bibfield  {author} {\bibinfo {author} {\bibfnamefont {J.~B.}\ \bibnamefont
  {Amado}}, \bibinfo {author} {\bibfnamefont {B.~C.}\ \bibnamefont {da~Cunha}},
  \ and\ \bibinfo {author} {\bibfnamefont {E.}~\bibnamefont {Pallante}},\
  }\href@noop {} {\  (\bibinfo {year} {2021})},\ \Eprint
  {http://arxiv.org/abs/2110.08349} {arXiv:2110.08349 [hep-th]} \BibitemShut
  {NoStop}%
\bibitem [{\citenamefont {da~Cunha}\ and\ \citenamefont
  {Cavalcante}(2021)}]{daCunha:2021jkm}%
  \BibitemOpen
  \bibfield  {author} {\bibinfo {author} {\bibfnamefont {B.~C.}\ \bibnamefont
  {da~Cunha}}\ and\ \bibinfo {author} {\bibfnamefont {J.~a.~P.}\ \bibnamefont
  {Cavalcante}},\ }\href {\doibase 10.1103/PhysRevD.104.084051} {\bibfield
  {journal} {\bibinfo  {journal} {Phys. Rev. D}\ }\textbf {\bibinfo {volume}
  {104}},\ \bibinfo {pages} {084051} (\bibinfo {year} {2021})},\ \Eprint
  {http://arxiv.org/abs/2105.08790} {arXiv:2105.08790 [hep-th]} \BibitemShut
  {NoStop}%
\bibitem [{\citenamefont {Amado}, \citenamefont {Carneiro~da Cunha},\ and\
  \citenamefont {Pallante}(2017)}]{Amado:2017kao}%
  \BibitemOpen
  \bibfield  {author} {\bibinfo {author} {\bibfnamefont {J.~B.}\ \bibnamefont
  {Amado}}, \bibinfo {author} {\bibfnamefont {B.}~\bibnamefont {Carneiro~da
  Cunha}}, \ and\ \bibinfo {author} {\bibfnamefont {E.}~\bibnamefont
  {Pallante}},\ }\href {\doibase 10.1007/JHEP08(2017)094} {\bibfield  {journal}
  {\bibinfo  {journal} {JHEP}\ }\textbf {\bibinfo {volume} {08}},\ \bibinfo
  {pages} {094} (\bibinfo {year} {2017})},\ \Eprint
  {http://arxiv.org/abs/1702.01016} {arXiv:1702.01016 [hep-th]} \BibitemShut
  {NoStop}%
\bibitem [{\citenamefont {Gibbons}, \citenamefont {Perry},\ and\ \citenamefont
  {Pope}(2005)}]{Gibbons:2004ai}%
  \BibitemOpen
  \bibfield  {author} {\bibinfo {author} {\bibfnamefont {G.~W.}\ \bibnamefont
  {Gibbons}}, \bibinfo {author} {\bibfnamefont {M.~J.}\ \bibnamefont {Perry}},
  \ and\ \bibinfo {author} {\bibfnamefont {C.~N.}\ \bibnamefont {Pope}},\
  }\href {\doibase 10.1088/0264-9381/22/9/002} {\bibfield  {journal} {\bibinfo
  {journal} {Class. Quant. Grav.}\ }\textbf {\bibinfo {volume} {22}},\ \bibinfo
  {pages} {1503} (\bibinfo {year} {2005})},\ \Eprint
  {http://arxiv.org/abs/hep-th/0408217} {arXiv:hep-th/0408217} \BibitemShut
  {NoStop}%
\bibitem [{\citenamefont {Hollands}, \citenamefont {Ishibashi},\ and\
  \citenamefont {Marolf}(2005)}]{Hollands:2005wt}%
  \BibitemOpen
  \bibfield  {author} {\bibinfo {author} {\bibfnamefont {S.}~\bibnamefont
  {Hollands}}, \bibinfo {author} {\bibfnamefont {A.}~\bibnamefont {Ishibashi}},
  \ and\ \bibinfo {author} {\bibfnamefont {D.}~\bibnamefont {Marolf}},\ }\href
  {\doibase 10.1088/0264-9381/22/14/004} {\bibfield  {journal} {\bibinfo
  {journal} {Class. Quant. Grav.}\ }\textbf {\bibinfo {volume} {22}},\ \bibinfo
  {pages} {2881} (\bibinfo {year} {2005})},\ \Eprint
  {http://arxiv.org/abs/hep-th/0503045} {arXiv:hep-th/0503045} \BibitemShut
  {NoStop}%
\bibitem [{\citenamefont {Olea}(2007)}]{Olea:2006vd}%
  \BibitemOpen
  \bibfield  {author} {\bibinfo {author} {\bibfnamefont {R.}~\bibnamefont
  {Olea}},\ }\href {\doibase 10.1088/1126-6708/2007/04/073} {\bibfield
  {journal} {\bibinfo  {journal} {JHEP}\ }\textbf {\bibinfo {volume} {04}},\
  \bibinfo {pages} {073} (\bibinfo {year} {2007})},\ \Eprint
  {http://arxiv.org/abs/hep-th/0610230} {arXiv:hep-th/0610230} \BibitemShut
  {NoStop}%
\bibitem [{\citenamefont {Motl}(2003)}]{Motl:2002hd}%
  \BibitemOpen
  \bibfield  {author} {\bibinfo {author} {\bibfnamefont {L.}~\bibnamefont
  {Motl}},\ }\href {\doibase 10.4310/ATMP.2002.v6.n6.a3} {\bibfield  {journal}
  {\bibinfo  {journal} {Adv. Theor. Math. Phys.}\ }\textbf {\bibinfo {volume}
  {6}},\ \bibinfo {pages} {1135} (\bibinfo {year} {2003})},\ \Eprint
  {http://arxiv.org/abs/gr-qc/0212096} {arXiv:gr-qc/0212096} \BibitemShut
  {NoStop}%
\bibitem [{\citenamefont {Berti}\ \emph {et~al.}(2003)\citenamefont {Berti},
  \citenamefont {Cardoso}, \citenamefont {Kokkotas},\ and\ \citenamefont
  {Onozawa}}]{Berti:2003jh}%
  \BibitemOpen
  \bibfield  {author} {\bibinfo {author} {\bibfnamefont {E.}~\bibnamefont
  {Berti}}, \bibinfo {author} {\bibfnamefont {V.}~\bibnamefont {Cardoso}},
  \bibinfo {author} {\bibfnamefont {K.~D.}\ \bibnamefont {Kokkotas}}, \ and\
  \bibinfo {author} {\bibfnamefont {H.}~\bibnamefont {Onozawa}},\ }\href
  {\doibase 10.1103/PhysRevD.68.124018} {\bibfield  {journal} {\bibinfo
  {journal} {Phys. Rev. D}\ }\textbf {\bibinfo {volume} {68}},\ \bibinfo
  {pages} {124018} (\bibinfo {year} {2003})},\ \Eprint
  {http://arxiv.org/abs/hep-th/0307013} {arXiv:hep-th/0307013} \BibitemShut
  {NoStop}%
\bibitem [{\citenamefont {Berti}, \citenamefont {Cardoso},\ and\ \citenamefont
  {Yoshida}(2004)}]{Berti:2004um}%
  \BibitemOpen
  \bibfield  {author} {\bibinfo {author} {\bibfnamefont {E.}~\bibnamefont
  {Berti}}, \bibinfo {author} {\bibfnamefont {V.}~\bibnamefont {Cardoso}}, \
  and\ \bibinfo {author} {\bibfnamefont {S.}~\bibnamefont {Yoshida}},\ }\href
  {\doibase 10.1103/PhysRevD.69.124018} {\bibfield  {journal} {\bibinfo
  {journal} {Phys. Rev. D}\ }\textbf {\bibinfo {volume} {69}},\ \bibinfo
  {pages} {124018} (\bibinfo {year} {2004})},\ \Eprint
  {http://arxiv.org/abs/gr-qc/0401052} {arXiv:gr-qc/0401052} \BibitemShut
  {NoStop}%
\bibitem [{\citenamefont {Klemm}(2014)}]{Klemm:2014rda}%
  \BibitemOpen
  \bibfield  {author} {\bibinfo {author} {\bibfnamefont {D.}~\bibnamefont
  {Klemm}},\ }\href {\doibase 10.1103/PhysRevD.89.084007} {\bibfield  {journal}
  {\bibinfo  {journal} {Phys. Rev. D}\ }\textbf {\bibinfo {volume} {89}},\
  \bibinfo {pages} {084007} (\bibinfo {year} {2014})},\ \Eprint
  {http://arxiv.org/abs/1401.3107} {arXiv:1401.3107 [hep-th]} \BibitemShut
  {NoStop}%
\bibitem [{\citenamefont {Hennigar}, \citenamefont {Kubiz\v{n}\'ak},\ and\
  \citenamefont {Mann}(2015)}]{Hennigar:2014cfa}%
  \BibitemOpen
  \bibfield  {author} {\bibinfo {author} {\bibfnamefont {R.~A.}\ \bibnamefont
  {Hennigar}}, \bibinfo {author} {\bibfnamefont {D.}~\bibnamefont
  {Kubiz\v{n}\'ak}}, \ and\ \bibinfo {author} {\bibfnamefont {R.~B.}\
  \bibnamefont {Mann}},\ }\href {\doibase 10.1103/PhysRevLett.115.031101}
  {\bibfield  {journal} {\bibinfo  {journal} {Phys. Rev. Lett.}\ }\textbf
  {\bibinfo {volume} {115}},\ \bibinfo {pages} {031101} (\bibinfo {year}
  {2015})},\ \Eprint {http://arxiv.org/abs/1411.4309} {arXiv:1411.4309
  [hep-th]} \BibitemShut {NoStop}%
\bibitem [{\citenamefont {Hennigar}\ \emph {et~al.}(2015)\citenamefont
  {Hennigar}, \citenamefont {Kubiz\v{n}\'ak}, \citenamefont {Mann},\ and\
  \citenamefont {Musoke}}]{Hennigar:2015cja}%
  \BibitemOpen
  \bibfield  {author} {\bibinfo {author} {\bibfnamefont {R.~A.}\ \bibnamefont
  {Hennigar}}, \bibinfo {author} {\bibfnamefont {D.}~\bibnamefont
  {Kubiz\v{n}\'ak}}, \bibinfo {author} {\bibfnamefont {R.~B.}\ \bibnamefont
  {Mann}}, \ and\ \bibinfo {author} {\bibfnamefont {N.}~\bibnamefont
  {Musoke}},\ }\href {\doibase 10.1007/JHEP06(2015)096} {\bibfield  {journal}
  {\bibinfo  {journal} {JHEP}\ }\textbf {\bibinfo {volume} {06}},\ \bibinfo
  {pages} {096} (\bibinfo {year} {2015})},\ \Eprint
  {http://arxiv.org/abs/1504.07529} {arXiv:1504.07529 [hep-th]} \BibitemShut
  {NoStop}%
\end{thebibliography}
\end{document}